\documentclass[fleqn,10pt]{wlscirep}
\usepackage[utf8]{inputenc}
\usepackage[T1]{fontenc}
\usepackage{bm}

\usepackage[style=numeric-comp,backend=biber,sorting=none]{biblatex}
\usepackage{xcolor,pifont} % for table check mark coloring
\usepackage{multirow} %for table
\usepackage{amsmath} %for table text use
\usepackage{graphicx}
\usepackage{wrapfig} %figure right/left aligned on page
\graphicspath{{images/}}
\usepackage{float}
\usepackage{comment}
\usepackage{blindtext}
\usepackage{array}    % für erweiterte Tabellenfunktionen
\usepackage{subfiles}
\DeclareUnicodeCharacter{202F}{FIX ME!!!!}
\usepackage{dirtytalk} %easier quotation marks, command: \say
\usepackage[british,UKenglish]{babel}
\DeclareFieldFormat{labelnumberwidth}{\mkbibbrackets{#1}}

\usepackage{verbatim}

\title{More than Carbon: Cradle-to-Grave environmental impacts of GenAI training on the Nvidia A100 GPU}
\author[1*]{Sophia Falk}
\author[2,3**]{David Ekchajzer}
\author[4]{Thibault Pirson}
\author[5]{Etienne Lees-Perasso}
\author[4]{Augustin Wattiez}
\author[6]{Lisa Biber-Freudenberger}
\author[7]{Sasha Luccioni}
\author[1]{Aimee van Wynsberghe}

\affil[1]{Sustainable AI Lab, Institute for Science and Ethics, Bonn University, Germany}
\affil[*]{corresponding author: falk@iwe.uni-bonn.de}
\affil[**]{corresponding author: da.ekchajzer@hubblo.org}
\affil[2]{Univ Evry, IMT-BS, LITEM, Université Paris-Saclay, Paris, France}
\affil[3]{Hubblo, France}
\affil[4]{Electronic Circuits and Systems Group, ICTEAM, UCLouvain, Louvain-la-Neuve, Belgium}
\affil[5]{TIDE}
\affil[6]{Center for Development Research, Bonn University, Germany}
\affil[7]{Hugging Face}

\begin{document}
\begin{abstract}
The rapid expansion of AI has intensified concerns about its environmental sustainability. Current assessments focus on operational carbon emissions using secondary data, overlooking impacts in other life cycle stages. This study presents a comprehensive multi-criteria life cycle assessment of AI training, examining 16 environmental impact categories using primary data from the Nvidia A100 GPU. Results for GPT-4 training show the use phase dominates 10 categories, contributing 96\% to climate change and fossil fuel depletion. Manufacturing dominates 6 categories, including human toxicity (94\%) and freshwater eutrophication (81\%). The GPU chip is the largest contributor in 10 categories, particularly climate change (81\%) and fossil resource use (80\%). While primary data produces modest changes in carbon estimates, substantial variations emerge elsewhere, e.g. minerals and metals depletion increases by 33\%. This analysis expands Sustainable AI discourse beyond carbon emissions, challenging current sustainability narratives.
\end{abstract}
\maketitle
\section*{Introduction}

`We need more data on AI!' - a statement as ironic as it is urgent. Despite being the most data-dependent technology in history, comprehensive data on AI’s environmental impacts across multiple categories remains remarkably incomplete.\\
As AI's existence depends entirely on the availability of powerful computational infrastructure, understanding the training hardware is crucial. Among the various hardware components that enable AI development, GPUs have emerged as the primary workhorses for model training, making their environmental assessment particularly important. 
Specifically, the release of the Nvidia A100 GPU in 2020 marks a cornerstone of AI training development, becoming one of the most widely deployed GPUs for AI training. 
The A100 is widely credited with revolutionizing deep learning efficiency \cite{MediumA100} by delivering nearly 2.5 times greater computational power and enhanced memory capacity compared to the previous GPU generation. 
This hardware advancement established a new benchmark for massive AI workloads, enabling the breakthrough model GPT-3 that redefined possibilities in natural language processing \cite{MediumA100}. 
The transformative impact of this technological advancement is reflected in Nvidia's record-breaking revenues, which showed 125.85\% growth from 2023 to 2024 \cite{NVIDIAfiscal2023,NVIDIAfiscal2024} - a direct consequence of unprecedented demand for AI-capable hardware. The availability of such hardware has, in turn, enabled rapid growth in AI model complexity from billions to trillions of parameters driving many of the recent advances in performance. A prominent example is GPT-4, estimated to comprise 1.8 trillion parameters \cite{GPT4_leaked}, whose scale alone necessitates exponential increases in computational power and consequently greater resource consumption \cite{debus2023reportingElectr, guo2025deepseek-R1}.\\
The combination of growing models and widespread AI adoption has fueled an unprecedented need for cutting-edge data centers equipped with state-of-the-art GPUs, the backbone of modern AI systems. 
Given the specialized nature and relatively short operational lifespan of these components \cite{Titan2020, meta2024llama3, karydopoulos2025heatimpact}, this fast-paced hardware expansion cycle raises important environmental concerns that extend beyond operational energy consumption. Nonetheless, current environmental assessments predominantly focus on energy consumption during AI model training, with carbon emissions varying based on model size, data center efficiency, and location-based grid carbon intensity \cite{strubell2019, lacoste2019quantifying, dodge2022measuring}. 
This carbon-centric approach, reinforced by existing literature and regulatory frameworks that prioritize energy consumption, has intensified attention on data center energy efficiency while overlooking broader environmental impacts arising from the manufacturing and disposal of specialized hardware. Thus, similar to other sectors, environmental assessments of AI remain constrained by a `carbon tunnel vision': a narrow focus on carbon emissions that overlooks the wider spectrum of environmental and social impacts \cite{jan_konietzko_moving_2022}. 
Yet, a single-criterion, carbon-centric approach that omits critical life cycle stages, such as raw material extraction and end-of-life processing, is insufficient for assessing the sustainability of AI. Environmental sustainability extends beyond climate change and requires consideration of potential burden shifting across different impact categories.
A comprehensive assessment encompassing material extraction, processing, manufacturing, and end-of-life stages is therefore essential for understanding AI’s complete environmental footprint.\\
The few publications offering multi-stage and multi-criteria approaches rely predominantly on secondary data sources and proxy estimates rather than primary data collection, limiting the accuracy and specificity of environmental impact assessments (for a detailed review, see the next Section). 
%%%%%%%%%%%%%%%%%%%%%%%%%%%%%%% BACKGROUND %%%%%%%%%%%%%%%%%%%%%%%%%%%%%%
Similar data limitations affect the operational phase, where AI can be either trained or used for inference. Although recent research has begun quantifying AI inference-related energy consumption, estimates vary widely with model size, input length, and prompt complexity \cite{luccioni2024inference}. Moreover, declining industry transparency around AI model specifications contributes to the spread of misinformation and contestation of published consumption estimates \cite{luccioni2025misinformation}.\\ 
Given these pervasive data limitations and transparency challenges across both embodied and operational stages, this study addresses the critical need for comprehensive, openly available primary data to close the fundamental gap in embodied impacts of AI hardware.
To capture AI's complete environmental footprint, we conduct the first comprehensive multi-criteria LCA across all life cycle stages (cradle-to-grave) of the Nvidia A100 SXM GPU based on primary data generated through a two-step methodology. 
First, we methodically disassembled the GPU into component groups through a detailed teardown analysis. Second, we performed a multi-element composition analysis to determine the material composition of each component group. This compositional analysis enables more precise component-level modeling and improves the accuracy of life cycle impacts assessment across the 16 environmental categories.
%%%
All primary datasets from the teardown and elemental analysis are openly available to enhance transparency and reproducibility\footnote{Find the supplementary materials here: \href{https://github.com/sophia-falk/more-than-carbon}{github.com/sophia-falk/more-than-carbon}}, thereby addressing the lack of accessible, high-quality data that has hindered progress in sustainability assessments in the field.\\
This study provides two key contributions to the field of sustainable AI.
First, it establishes a comprehensive foundation of primary data for AI hardware environmental assessment, addressing the critical lack of component-specific material inventories that has limited previous studies to secondary data and proxy estimates. 
Second, it expand the Sustainable AI discourse beyond its narrow focus on operational carbon emissions to include a comprehensive multi-indicator ecological footprint spanning the complete life cycle.
Together, these contributions provide valuable insights for policymakers developing AI governance frameworks, industry leaders implementing sustainable practices, and researchers advancing environmental assessment methodologies, supporting the development of effective strategies to mitigate AI's environmental impacts while ensuring sustainable technological progress.
 
\section*{Literature Review and Background}

This section presents a literature review of studies establishing theoretical and methodological foundations for AI environmental sustainability assessments. The review concentrates on research considering comparable AI infrastructure and LCA frameworks for evaluating ML/AI training processes. This targeted approach enables focused examination of the most relevant methodological precedents and empirical findings in this area.\\
\\
While specialized GPUs provide computational power for AI training and inference, they operate within broader systems including servers, racks, high-speed networking equipment, cooling systems, and power supplies. Research on the environmental assessment of AI infrastructures can be conducted at multiple levels using different system boundaries ranging from individual AI chips to GPU cards to comprehensive evaluations of GPU-Server combinations, IT rooms and data center facilities. 
The reviewed studies span different infrastructure boundaries and life cycle stages, ranging from the use phase, to including manufacturing stage, to cradle-to-grave approaches (as illustrated in Fig.\ref{fig:GPU2FU} (a+d)). Cradle-to-gate stages encompass the combined life cycle phases of resource extraction, processing, and
manufacturing, while cradle-to-grave extends this scope to include the use phase and end-of-life.
Assessment criteria range from single-indicator approaches focused on carbon emissions to multi-criteria evaluations that can incorporate up to 16 environmental impact categories. Table \ref{tab:backgorund_lit} provides a comprehensive overview of these studies, summarizing their AI infrastructure boundary, functional unit, data source, scope, and impact categories to contextualize our study within the existing literature.\\
The evolution of AI environmental assessment reflects this diversity in scope and methodology. Early AI sustainability research concentrated exclusively on operational carbon emissions during model training. Strubell et al. (2019) established foundational quantification methods \cite{strubell2019}, with subsequent studies adopting increasingly sophisticated methodologies. For instance, Lacoste et. al (2019) enhance precision by incorporating geographical energy mix factors for carbon intensity estimation \cite{lacoste2019quantifying}, while Dodge et al. (2022) employ location-based and time-specific marginal emissions data per energy unit during model training \cite{dodge2022measuring}.\\
Subsequent research expands to include embodied emissions from AI infrastructure manufacturing. Gupta et al. (2021) integrate insights from consumer electronics carbon footprint studies (smartphones, smart speakers, laptops, tablets) to comprehend and address the embodied carbon impact of data center hardware, using industry-reported sustainability data. They suggest that a significant proportion of the carbon footprint is attributable to AI hardware manufacturing \cite{gupta2021chasing}.\\
%
%Wu 2022facilitates
Wu et al. (2022) examine both use phase and manufacturing related carbon footprints across Meta's industry-scale ML infrastructure using Apple's 28-core CPU with dual AMD Radeon GPU as a proxy reference due to limited availability of LCA data for actual AI-specific GPUs and accelerators. Their findings demonstrate that embodied carbon represents approximately 50\% of the total environmental impact for large-scale ML tasks when renewable energy sources are considered for operational emissions \cite{wu2022sustainable}.
%
% Li 2025
Li et al. (2025) incorporate embodied carbon considerations for AI infrastructure on the data center-level. The study relies entirely on secondary data sources and proxy estimates from industry reports, most notably using the Dell R740 LCA as a universal baseline and scaling it across different AI system configurations through area-based and power-based scaling factors \cite{Li2025ecoserve} - a significant methodological limitation compared to studies with direct component-specific LCA data. Their assessment covers cradle-to-gate and the use phase for a single-indicator expressed in carbon emissions. 
In line with findings from \cite{gupta2021chasing} and \cite{wu2022sustainable}, Li et al. (2025) reveal that manufacturing host processing systems (servers, memory, storage) dominate embodied carbon contributions while GPUs dominate operational carbon, suggesting that optimizing only GPU efficiency misses significant opportunities for carbon reduction \cite{Li2025ecoserve}. The studies conducted by Gupta et al. (2021), Wu et al. (2022), and Li et al. (2025) highlight the necessity to account for additional life cycle stages beyond the operational phase. Nonetheless, the three studies under consideration are dependent on secondary and proxy data for the embodied emissions calculation. This underscores the critical necessity for the availability of AI-specific data for LCA studies.\\
\begin{table}
    \centering
    \caption{Summary of reviewed literature on AI environmental sustainability assessment, showing infrastructure system boundaries (from individual chips to complete data centers), functional units (AI training vs. inference focus), life cycle inventory (LCI) data sources (primary vs. secondary), and assessment scope (life cycle stages and environmental indicators covered).}
    \renewcommand{\arraystretch}{1.1} % Adjust row spacing
    \newcommand*\colourcheck[1]{%
  \expandafter\newcommand\csname #1check\endcsname{\textcolor{#1}{\ding{52}}}%
}
\newcommand{\xmark}{\ding{55}}%
    \colourcheck{green}
    \resizebox{\textwidth}{!}{%
    \Large  
    \begin{tabular}{l|llc|c c c c|c c} %\hline

    \toprule[1.5pt]
         \Large\textbf{Publication} & \Large\textbf{AI Infrastructure}& \Large\textbf{Functional}& \Large\textbf{Data source} & \Large\textbf{Resource extraction \&}& \Large\textbf{Manu-} & \Large\textbf{Use}& \Large\textbf{End-of}& \Large\textbf{Single criterion:} & \Large\textbf{Multi-} \\
         & \Large\textbf{ Boundary}& \Large\textbf{Unit}& \Large\textbf{(hardware LCI)}& \Large\textbf{Processing}& \Large\textbf{facturing} & \Large\textbf{phase}& \Large\textbf{-life}& \Large\textbf{Carbon emissions}& \Large\textbf{Criteria} \\ %\hline
    \toprule[1.5pt]
         
         Strubell et al. (2019) \cite{strubell2019} &  GPU Card&  Training NLP model&  \xmark&  \xmark& \xmark & \greencheck& \xmark & \greencheck& \xmark \\ %\hline
         Lacoste et al. (2019) \cite{lacoste2019quantifying} &  GPU Card&  Training ML model&  \xmark&  \xmark&  \xmark&  \greencheck&  \xmark&  \greencheck& \xmark\\ %\hline
         Dodge et al. (2022) \cite{dodge2022measuring} &  GPU Card&  single AI cloud instance&  \xmark&  \xmark&  \xmark&  \greencheck&  \xmark&  \greencheck& \xmark\\ %\hline
         Gupta et al. (2021)\cite{gupta2021chasing} &  Server&  not defined&  secondary \& proxy data&  \greencheck&  \greencheck&  \greencheck&  \xmark&  \greencheck& \xmark\\ %\hline
         Wu et al. (2022) \cite{wu2022sustainable}&  Server &  Training ML model&  secondary \& proxy data&  \greencheck&  \greencheck&  \greencheck&  \xmark&  \greencheck& \xmark\\ %\hline
         Li et al (2025) \cite{Li2025ecoserve} &  Data Center&  per AI inference task &  secondary \& proxy data&  \greencheck&  \greencheck&  \greencheck&  \xmark&  \greencheck& \xmark\\ %\hline
         De Vries et al. (2025) \cite{greenpeace2025} &  GPU Chip&  not applicable&  secondary data&  \xmark&  \greencheck&  \xmark&  \xmark&  \greencheck& \xmark\\ %\hline
         NVIDIA (2025) \cite{NvidiaH100PCF} &  Server &  not applicable&  primary data &  \greencheck&  \greencheck&  \xmark&  \xmark&  \greencheck& \xmark\\ %\hline
         Luccioni et al. (2023) \cite{luccioni2023} &  Data Center &  Training AI model&  secondary \& proxy data&  \greencheck&  \greencheck&  \greencheck&  \xmark&  \greencheck& \xmark\\ %\hline
          Berthelot et al. (2024) \cite{berthelot2024estimating} &  Server &  AI /ML training &  secondary \& proxy data&  \greencheck&  \greencheck&  \greencheck&  \xmark&  \xmark& 3 \greencheck\\ %\hline
         Morand et al. (2024) \cite{Morand2024MLCA} &  Server &  AI /ML training instance&  secondary \& proxy data&  \greencheck&  \greencheck&  \greencheck&  \xmark&  \xmark& 3 \greencheck\\ %\hline
         Desroches et al. (2025) \cite{desroches2025exploringsustainablescalingai} &  IT room&  Training AI model \& AI inference&  secondary data&  \greencheck&  \greencheck&  \greencheck&  \xmark&  \xmark& 5 \greencheck\\ %\hline
         Schneider et al. (2025) \cite{schneider2025lifecycleemissionsaihardware} &  Server &   Training AI model \& AI inference&  primary data &  \greencheck&  \greencheck&  \greencheck&  \greencheck&  \greencheck& \xmark\\ \hline
         %\hline
     %\toprule[1.5pt]
         \textbf{This study}&  GPU Card&  Training AI model&   primary data &  \greencheck&  \greencheck&  \greencheck&  \greencheck&  \xmark& 16 \greencheck\\ %\hline
     \bottomrule[1.5pt]
    \end{tabular}%
}
    \label{tab:backgorund_lit}
\end{table}
\noindent
De Vries (2025) specifically examines GPU chip manufacturing emissions using secondary data from industry reports, manufacturing specifications and market analysis reports to estimate production volumes and technical specifications \cite{greenpeace2025}. However, the study's reliance on secondary sources and single-indicator carbon assessment limits accuracy compared to primary data collection methods from teardown and elemental analysis.
%NVIDIA 
The Nvidia (2025) product carbon footprint report is the first industry-reported cradle-to-gate assessment for an AI server using a combination of primary supplier data of critical components including material composition and production energy consumption data \cite{NvidiaH100PCF}. However, limited transparency in terms of hardware specifications, modeling assumptions, and calculation methodologies, restricts reproducibility and independent verification.
%Luccioni
The cradle-to-gate plus operational phase approach by Luccioni et al. (2023) establishes precedent for comprehensive AI model assessment. The authors assess BLOOM's training carbon footprint focusing on the data center-level entirely relying on secondary and proxy data sources from a comparable HPE ProLiant server and estimate GPU embodied emissions due to limited manufacturer disclosure \cite{luccioni2023}.\\
%Berthelot and Morand
Berthelot et al. (2024) conduct the first multi-criteria LCA expressing impacts results in climate change, resource depletion, and primary energy demand, followed by Morand et al. (2024). Both studies, however, depend on secondary and proxy life cycle inventories due to the limited availability  of primary data for specialized hardware such as GPUs. 
Desroches et al. (2025) follow a cradle-to-gate plus use-phase approach at the IT room level. While this multi-criteria study evaluates five environmental indicator categories - climate change, water use, abiotic resource depletion, primary energy consumption, and final energy consumption - it relies on secondary and proxy data sources  \cite{desroches2025exploringsustainablescalingai}.
%Schneider
Finally, Schneider et al. (2025) conduct the first comprehensive cradle-to-grave LCA of AI hardware, focusing on Google's Tensor Processing Units (TPUs) and their complete computing infrastructure at the data center level. The study employs high-quality primary data, supplemented by supplier-provided data for thermal solutions and established LCA databases for standard components, representing the most comprehensive primary data collection for AI manufacturing emissions published to date \cite{schneider2025lifecycleemissionsaihardware}. However, the assessment is limited to a single-indicator scope and Google's TPUs, which limits generalizable to other AI hardware use cases. \\
The reviewed literature reveals significant methodological limitations that constrain comprehensive understanding of AI's environmental impacts. Ten of twelve studies employ single-criterion carbon-centric approaches, with only Morand et al. (2024) and Desroches et al. (2025) adopting multi-criteria assessments covering 3 and 5 impact categories respectively. Most existing analyses omit critical life cycle stages. The few publications offering more life cycle stages coverage (multi-stage and multi-criteria) rely predominantly on secondary data sources and proxy estimates rather than primary data collection, limiting accuracy and specificity of environmental impacts assessment. Furthermore, to our knowledge, none of these studies have performed an elemental analysis of electronic equipment in their LCA methodology.\\
This study addresses these critical gaps by presenting the first cradle-to-grave multi-criteria LCA covering 16 impact categories of AI hardware based on comprehensive primary data collection. Through detailed physical teardown analysis and elemental composition characterization of GPU components, we provide precise material inventories and manufacturing specifications that enable a more accurate environmental impacts assessment across multiple indicators. 
Furthermore, we make all primary datasets openly available to enable reproducibility, support further research, and enhance transparency in AI environmental impacts assessment - addressing the lack of accessible, high-quality data that has hindered progress in this field.
\section*{Method}

This section describes the goal and scope and the modeling principles of this study, following ISO 14040:2006 \cite{iso14040:2006} and ISO 14044:2006 \cite{iso14044:2006} standards, ensuring methodological rigor and transparency. 
An attributional modeling framework is applied to quantify the environmental burdens and resource consumption across multiple impact categories that are specifically attributable to AI hardware production and use. The selection of this approach over a consequential framework is driven by both methodological and practical considerations. The focus of this study is to understand current environmental impacts of training AI models rather than assessing changes in demand for AI over time or market response to increased AI model development. Moreover, given that the secondary objective of this research is to raise awareness about AI hardware's environmental impacts among non-LCA practitioners, including policymakers, industry stakeholders, and the broader scientific community, the attributional framework's accessibility and interpretability is particularly well-suited. 

\subsection*{Goal and Scope}

\textit{Goal} The objective of this study is twofold. 
The primary goal is to provide LCA practitioners, scientists conducting research on the environmental footprint of information and communication technologies (ICT) and the AI research community with high-quality primary data and results for multi-criteria cradle-to-grave analysis of an Nvidia A100 SXM 40 GB GPU used for AI training and inference. This foundational data can be used to conduct in-depth AI-as-a-service related environmental impacts assessments. 
Our secondary goal addresses the growing public attention on AI's sustainability by serving as a resource for stakeholders (e.g. journalist) seeking to communicate the environmental impacts beyond climate change to a wider audience outside of scientific research through scientistic popularization.\\
For our primary goal of creating a detailed primary dataset, we assess the environmental impacts associated with training BLOOM. We select BLOOM because it has been extensively discussed in the scientific literature and represents an important case study in ongoing academic discussions about transparency in AI model development and environmental impact assessment. 
For our secondary goal of raising awareness within the Sustainable AI discourse, we evaluate the environmental impacts of training GPT-4. This widely recognized model was chosen to enhance communication of AI's environmental impacts to broader audiences beyond the academic community. 

\begin{table}[h]
    \centering
    \small % Sets font size to approximately 11pt
    \caption{Configuration 1 and 2 for the Functional Unit: The complete training of one X-parameter Y model using Z GPU hours on Nvidia A100 SXM 40GB in location A during year B, resulting in a production-ready model capable of providing inference as a service.}
    \renewcommand{\arraystretch}{1.1} % Adjust row spacing
    \begin{tabular*}{\textwidth}{@{\extracolsep{\fill}} p{2.25cm} @{\hspace{5pt}} p{3cm} @{\hspace{5pt}} p{2.5cm} @{\hspace{5pt}} p{3cm} @{\hspace{5pt}} p{2cm} @{\hspace{5pt}} p{1.25cm}}
        % \hline
       \toprule[1.5pt]
        \textbf{Model (X)} & \textbf{Domain (Y)} & \textbf{GPU hours (Z)} & \textbf{Hardware} & \textbf{Country (A)} & \textbf{Year (B)} \\
       \toprule[1.5pt]

        %\hline
        BLOOM (176B) & Large language model & 1,080,000 & A100 SMX 80GB & Paris, France & 2022 \\
         %\hline
  GPT- 4 (1.8T) & Multimodal model & 57,000,000 & A100 SMX 40/80GB  & Iowa, USA & 2023  \\
   %\hline
   \bottomrule[1.5pt]
    \end{tabular*}
    \label{tab:AI_models}
\end{table}

\begin{wrapfigure}{r}{0.5\textwidth}
    \centering
    \includegraphics[width=0.5\textwidth]{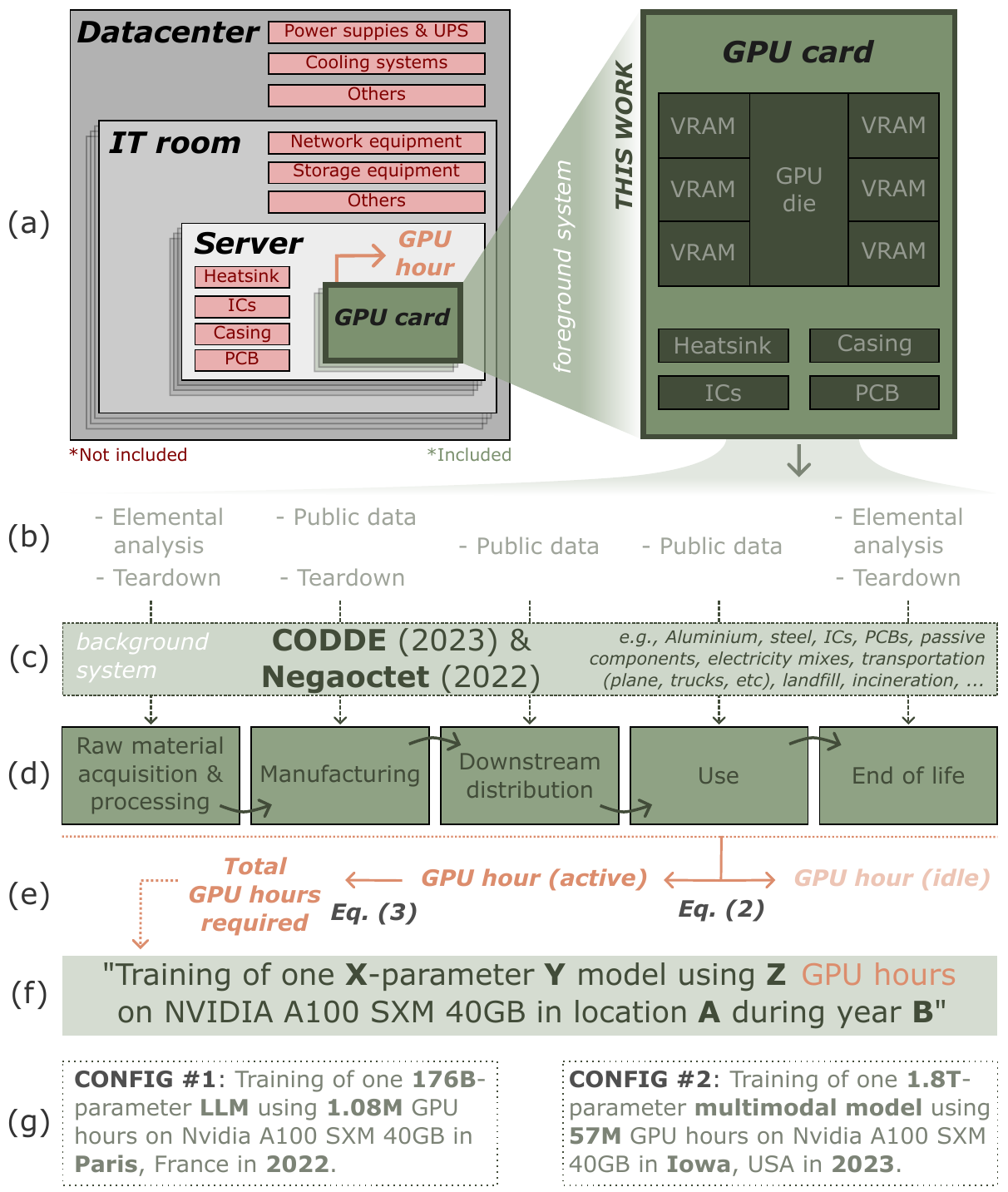}
    \caption{Modeling principles of this study. (a) Generic technology context for the product system and foreground system. (b) Data collection methodology. (c) Background system. (d) Life cycle stages. (e) Allocation for the reference flow. (f) Functional unit. (g) Two different configurations considered in this study to illustrate the model on real-life case studies (author illustration). }
    \label{fig:GPU2FU}
\end{wrapfigure}

\noindent
\textit{Scope} This study examines the multi-criteria environmental impacts from cradle-to-grave of training AI services on a high-performance computing (HPC) cluster providing GPU hours. GPU hours are a common metric to measure AI training duration and costs. 
From its release in 2020 until late 2023, the A100 was the most widely used hardware inside HPC clusters for AI training, playing a crucial role in the development of large-scale AI models. The Life Cycle Impacts Analysis (LCIA) method for the 16 selected criteria relies on the Product Environmental Footprint (PEF) method based on the EU recommendation 2013/179/EU \cite{EUreco2013/179/EU}, which aligns with ISO 14040/44 standards. Indicators are set for PEF EF 3.0 v2.0 (midpoint) \cite{EPLCA_DeveloperEF}. The modelling approach employed is illustrated in Figure \ref{fig:GPU2FU}.\\
\noindent
\textit{Functional unit} GenAI models do not always fulfill the same purpose, this also holds for BLOOM and GPT-4\footnote{unofficial sources estimate the training ranged from 54,000,000- 60,000,00 GPU hours \cite{GPT4_leaked}.}. While BLOOM is a large language model focused on text processing, GPT-4 is a multimodal model capable of processing both text and images. These functional differences make the definition of one functional unit (FU) challenging. Therefore, this study employs a parametric FU with the aim of capturing the essential training characteristics of both models. The FU is defined as: \\
The complete training of one \textbf{X}-parameter \textbf{Y} model using \textbf{Z} GPU hours on Nvidia A100 SXM 40GB in location \textbf{A} during year \textbf{B}, resulting in a production-ready model capable of providing inference as a service.\\
\\
We configure the FU for both analyzed models using the values provided in Table \ref{tab:AI_models} for each parameter X, Y, Z, A and B. Note, that this study does not constitute a comparative LCA, as BLOOM and GPT-4 provide different services and capabilities upon completion of training.\\
\\
\textit{Database} The foreground system of our life cycle model is mainly constructed with primary data collected through a thorough teardown and elemental analysis.  The disaggregated data is available in the supplementary material (SM), which is made available open source in accordance with the principles of transparency, reproducibility and re-usability. In addition, public data provided by Nvidia, Hugging Face and OpenAI is used, as well as hypotheses. The background system is based on the Negaoctet database for ICT-specific data (version 11/02/2022) and the CODDE database is used for generic processes (version CODDE-2023-02). The disclosure of specific background data is prohibited due to proprietary constraints. The LCIA for the graphic cards has been conducted in the EIME software.\\ 
\\
\textit{Modeling principles} In order to assess the impacts attributed to a specific training (as defined in the FU), the initial step involves the evaluation of the graphic card itself. The impacts for a single card are computed as intermediate results. In the context of this study, GPUs are regarded as providing a certain quantity of \textit{active GPU hours}. 
The latter is defined as the total duration for which a GPU is allocated to a specific training task over the course of its operational lifespan. The quantity of \textit{active GPU hours} provided by a single GPU is dependent on its \textit{lifespan} and \textit{utilization ratio}, as defined in equation \ref{eqn:gpu_per_hour}:
\begin{equation}
    \text{active\_GPU\_hours} ={\text{lifespan}_\text{{hours}}*\text{utilization\_ratio}}  
    \label{eqn:gpu_per_hour}
\end{equation}
\noindent
The impacts of the GPU are then distributed across the \textit{active GPU hours}, as shown in Equation \ref{eqn:impact_gpu_per_hour}. In situation where the  \textit{utilization ratio} is less than 100\%, this enables a comprehensive allocation of all impacts associated with the card, enabling the attribution of these impacts exclusively to useful GPU hours excluding idle hours.
\begin{equation}
    \text{impact}_\text{{\texttt{active\_GPU\_hour}}} =  \frac{\text{impact}_\texttt{{GPU}}}{\text{active\_GPU\_hours}}  
    \label{eqn:impact_gpu_per_hour}
\end{equation}\\
\textit{Active GPU hours} constitute our \textit{reference flow}, used to quantify the necessary amount of inputs for the fulfillment of the FU. The impacts of the training is then determined by multiplying the quantity of GPU hours required to train the model by the impacts of one \textit{active GPU hour} (see equation \ref{eqn:impact_FU}). The quantity of \textit{GPU hours} for each model (depending on FU configuration) are reported in Table \ref{tab:AI_models}.
\begin{equation}
    \text{impact}_\text{{\texttt{FU}}} =    \text{impact}_\text{{\texttt{active\_GPU\_hour}}} *   \text{GPU\_hours}_\text{{\texttt{training}}}
    \label{eqn:impact_FU}
\end{equation}
\\
\textit{Normalization}
In order to provide further support for the communication objective, the LCA results are normalized against a reference situation to eventually provide LCIA normalized results, enabling comparison across diverse environmental indicators that would otherwise have incompatible units and magnitudes. 
Among several normalization approaches available, we select the Planetary Boundary (PB)-aware normalization defined by Sala et al (2018) \cite{sala2018_PB_LC_development} as the most relevant for our study objectives. Contrarily to common reference scenario, the PB-aware reference defines annual values of individual impacts aligned with ecological limits, for each impact category. This normalization employs the widely recognized PB framework to enhance the interpretation of results for non-academic stakeholders, thereby facilitating more effective communication of AI's environmental significance in a global context.  
Other available LCIA normalization methods typically rely on historical reference scenarios, while the PB-aware normalization anchors our results to the ongoing debate on current planetary limits. Instead of expressing the impacts as a proportion of the actual or historical impacts per capita, this approach expresses the impacts as the proportion of ecological limits per capita considering planetary limits. 
To contextualize the LCA results within this framework, impacts are normalized to the world population of 2023 \cite{WorldBank2023Pop}, as this is the year that GPT-4 was trained in.
This representation allows all impact categories to be normalized on the same scale - one inhabitant per year budget. Hence, the normalized metrics express the environmental impacts of training an AI model as the proportion of the annual `environmental budget' per world capita of 2023. However, it is important to note that this per capita normalization does not indicate whether training an AI model is compatible with the global safe operating space (SOS). The results only express the impacts of training an AI model in the equivalence of PB-aware personal budgets. Determining the absolute sustainability of training AI through absolute environmental sustainability assessment (AESA) would require extra-steps such as down-scaling/allocation of carrying capacities and the interpretation of absolute sustainability ratios - for instance, scaling the PBs to our FU and comparing against total global AI training activities - which is far beyond the scope of this study.\\

\subsection*{Data collection}
A comprehensive data collection phase was conducted focusing on the cradle-to-gate stages. Further, we specify the downstream distribution assumptions, derive a use phase equation, and describe the end-of-life allocation approach employed to complete the life cycle inventory.

\subsubsection*{Cradle-to-Gate}
\begin{wrapfigure}{r}{0.5\textwidth}
    \centering
    \includegraphics[width=0.5\textwidth]{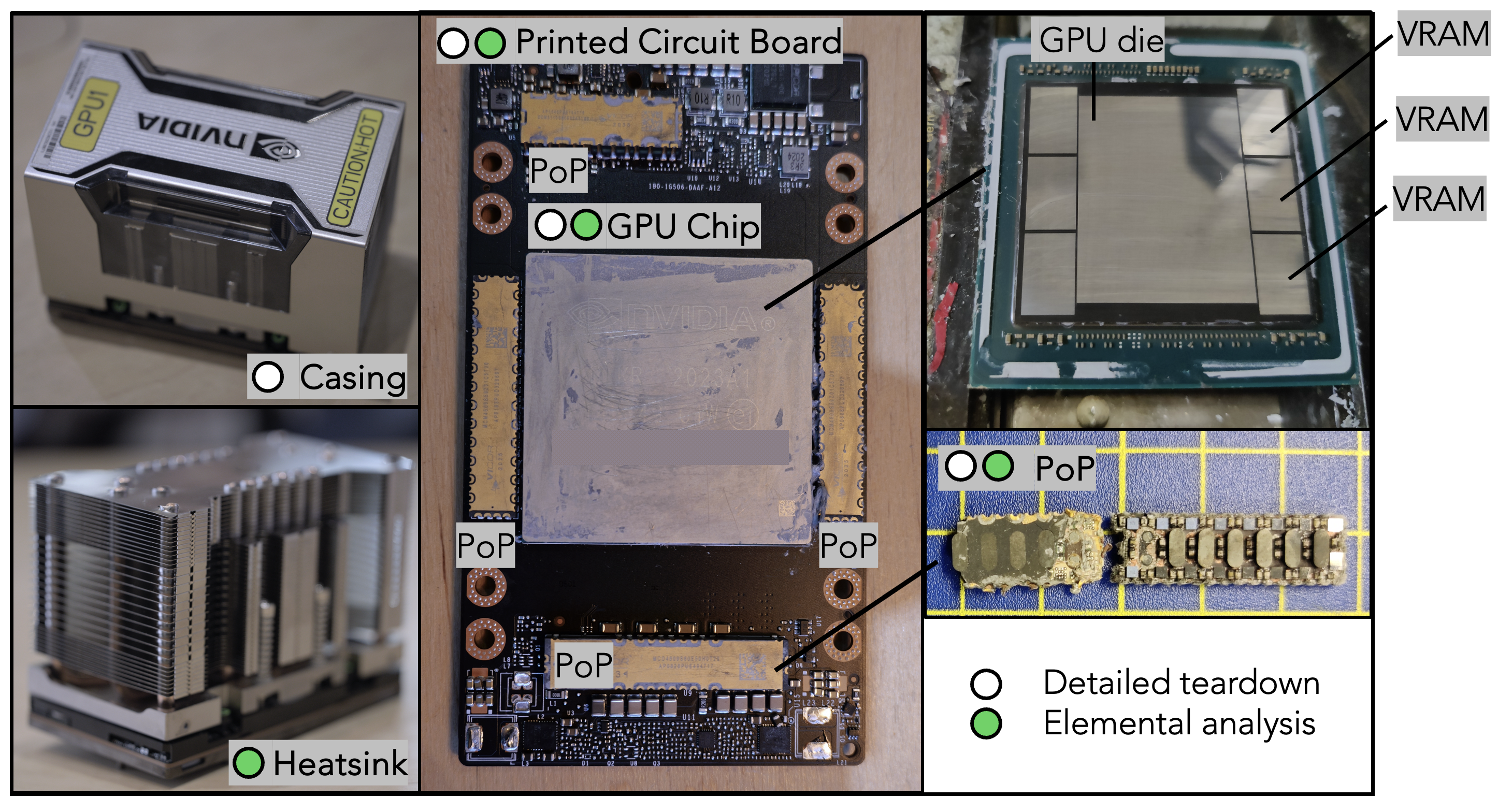}
    \caption{Annotated components of the Nvidia SXM A100 40 GB GPU showing which components were the subject of a detailed teardown analysis and which underwent an elemental analysis via ICP-OES (author pictures).}
    \label{fig:annotated components}
\end{wrapfigure}

The primary data collection process for the cradle-to-gate stages comprises two main steps. First, a teardown analysis is performed on the Nvidia A100 SXM GPU to create a bill of materials (BOM). This involves an in-depth examination of the main electronic components stacked on the PCB based on author's expertise in circuit design and LCA of ICT equipment. Secondly, an elemental composition of the individual GPU components is conducted using inductively coupled plasma optical emission spectroscopy (ICP-OES) to adapt the metals flows modeled based on the generic processes present in the background databases. More details can be found in the SM.
\begin{table}[H]
    \centering
    \small % Sets font size to approximately 11pt
    \caption{This overview shows which components were the subject of a detailed teardown analysis and which underwent a detailed elemental analysis via ICP-OES.}
    \renewcommand{\arraystretch}{1.0} % Tighter row spacing
    \newcommand*\colourcheck[1]{%
    \expandafter\newcommand\csname #1check\endcsname{\textcolor{#1}{\ding{52}}}%
    }
    \newcommand{\xmark}{\ding{55}}%
    \colourcheck{green}
    \begin{tabular}{p{4.5cm} >{\centering\arraybackslash}p{3cm} >{\centering\arraybackslash}p{3cm}}
        %\hline
        \toprule[1.5pt]
         \textbf{Components}&  \textbf{Detailed teardown}&\textbf{Elemental Analysis}\\
         %\hline
         \toprule[1.5pt]
         Casing&  \greencheck&\xmark\\ %\hline
         Heatsink& \greencheck&\greencheck\\ %\hline
         Printed Circuit Board&  \greencheck&\greencheck\\ %\hline
         GPU die + VRAM&  \greencheck&\greencheck\\ %\hline
         Power-on-Package (PoP)&   \greencheck&\greencheck\\
         %\hline
         \bottomrule[1.5pt]
    \end{tabular}
    \label{table:component_details}
\end{table}
\noindent
\textit{Detailed tear-down analysis}\\
The component inventory is constructed through a detailed teardown analysis of a Nvidia A100 SXM 40GB GPU. The product is first disassembled into five component groups: casing, heatsink, printed circuit board (PCB), Power-on-Package\footnote{The PoP is a highly specific piece of hardware, which is not found in common electronic equipments. Its main purpose includes scaling down and regulating voltage while significantly increasing the current to be delivered to the GPU (which can reach very high peaks of current, up to 600 to 1000A).} (PoP), and GPU chip (VRAM and GPU die) (see Fig. \ref{fig:annotated components}). The largest integrated circuits are additionally desoldered for further analysis. 
\noindent
Relevant input parameters such as the weight or the surface for each component are recorded. In particular, the die area is an essential input parameter for electronic component for properly scaling the LCI dataset in the Negaoctet database, as it is provided per cm$^2$ of die for most components. The dimensions of the semiconductor die inside the PoPs and largest integrated circuit are revealed by sinking components in a hot sulfuric acid bath, a technique known as chemical decapping \cite{pirson2021assessing}. It selectively removed the external packaging, exposing the die area inside. The die areas are then measured using a caliper, achieving a resolution of less than 0.01mm. In the specific case of PoPs, the internal structure contained several ICs. Such a component complexity is far from being captured in current LCA databases, even the ones with specific modeling of electronic components. Therefore, PoPs require extra assumptions to be modeled with best available proxies (see SM). The chemical decapping procedure is not necessary for the GPU die, as it uses a flip-chip package design where the semiconductor die area is only protected by a metallic shield, which is removed using a soldering gun.\\
\\
\textit{Elemental Analysis}\\
ICP-OES is an analytical technique that determines the qualitative and quantitative elemental composition of materials. The disassembled GPU components (see Figure \ref{fig:annotated components}) are subjected to controlled sample preparation before the ICP-OES analysis. In a first step, the samples (divided by component group)(see Table \ref{table:component_details}) are prepared by manually grinding them to a small particle size, followed by pyrolysis at 500°C to degrade all plastics, and then ground again to obtain a smaller particle size of $\leq$2mm. Three different mineralization procedures are employed to each sample using various acid combinations (HNO$_3$/H$_2$O$_2$, aqua regia, and tetrafluoroboric acid/nitric acid) to dissolve different metal groups, including difficult-to-detect elements like silicon. The resulting solutions are analyzed using ICP-OES with both qualitative and quantitative methods to determine metal concentrations in the product samples. More details on the sample preparation and analysis can be found in the SM. The results of the analysis are then used to adjust the initial modeling of the quantities of elements for each group of analyzed components. 
The elemental composition of the final product constitutes only a subset of the elements employed during the manufacturing of the GPU card. This can be attributed to the subtractive nature of certain manufacturing processes, such as the engraving of semiconductors. In order to take losses throughout the process into consideration, a loss ratio is applied to each element. The value of this ratio depends on the element's price. In cases where the element is more costly, the loss is regarded as being lower.
Where possible, each quantity is associated with the corresponding manufacturing process in the CODDE database. Otherwise, the elementary flow data is used, which ignores the impacts induced by the element's transformation processes, but retains those related to resource depletion. Only 11 out of 46 elements are affected by this approach.  
For the casing component, generic processes from the CODDE database are used since the elemental analysis method would burn off most of the casing material without providing meaningful material composition insights. The detailed BOM is provided in the SM.

\subsubsection*{Downstream Distribution}
We hypothesize that the GPU cards are transported by air from their assembly location (presumed to be Shenzen) to their final destination (Paris, France for BLOOM and Des Moines, Iowa for GPT-4). This assumption is justified by the highly competitive market for these components, which demands rapid delivery times. The weight of each card is 1.5 kilograms, with an additional 20\% added to account for packaging. We then apply an LCI data from the background database for cargo transportation by jet-powered aircraft.

\subsubsection*{Use phase}
In accordance with the defined scope, the environmental impacts during the use phase are solely dependent on the energy consumption of the GPUs. 
The power consumption of GPUs is contingent upon their utilization. Because AI training is a computationally highly intensive task, we use the maximum thermal design power (TDP) of  the A100 SXM (400 W) as the average power consumption per GPU card \cite{NvidiaA100Product}. This parameter will be subject to a Sensitivity Analysis (SA). When GPUs are not allocated to a training task, the equipment is estimated to consumes an idle power of 85W \cite{vamja2025partitioning_idle}.

\begin{equation}
    \text{electricity\_consumption}_\text{{\texttt{Wh}}} = ({\text{avg\_power}_\texttt{{W}}*\text{utilization\_ratio}}+\text{idle\_power}_\texttt{{W}}*(1-\text{utilization\_ratio}))*{\text{lifespan}_\texttt{{hours}}}
        \label{eqn:elec_conso_gpu}
\end{equation}
\noindent
The impacts of electricity consumption depend on the composition of the electricity mix where the data center is located. For each FU configuration, we use the impacts per kWh corresponding to the year and location of the training: 2022 in France for BLOOM \cite{luccioni2023} and 2023, Iowa, USA for GPT-4 \cite{microsoft_Iowa}. The impacts for the French electricity grid mix are taken from the CODDE database, while the impacts for the Iowa electricity mix are assessed based on the grid mix provided by the Iowa Utilities Commission \cite{Iowa2023}. The electrical mix considered during GPU chip manufacturing is Taiwan's energy grid taken from the CODDE database.

\subsubsection*{Training Infrastructure Parameters}
\textit{Utilization ratio} Large-scale data centers typically divide their infrastructure into distinct partitions to accommodate different types of workloads - most commonly training, inference and other deep learning tasks \cite{ye2024deep}. We investigate a GPU cluster that is used exclusively for training purposes in the training partition of a data center. We estimate a GPU \textit{utilization\_ratio} of 100\% for the training clusters. This assumption is justified by two key factors: first, training workloads are highly predictable compared to inference tasks, enabling effective resource planning; second, specialized AI training data centers deploy sophisticated workload schedulers specifically designed to minimize idle time and maximize utilization. While inference clusters may experience significant under-utilization due to unpredictable demand spikes, training partitions can sustain near-optimal utilization through proper workload management and exclusive GPU allocation methods used for large-scale foundational model training. This parameter will be subject to a SA.\\
\noindent
\textit{Lifespan} Based on extensive real-world data measurements in cluster computers \cite{Titan2020, meta2024llama3, karydopoulos2025heatimpact} (detailed in SM), we adopt an average three-year \textit{lifespan} for the A100 GPU in our study. This parameter will also be subject to a SA.

\subsubsection*{End-of-life}
The recycling and recovery of materials at the end-of-life (EoL) stage are considered using the formula and data developed by Ecosystem \cite{ecosystem}. 
The allocation approach chosen is the cut-off method at the ‘no benefit’ substitution point. This approach is based on the principle that recycling or energy recovery of EoL materials does not provide benefits in terms of substituting virgin materials or primary energy sources. This implies that no environmental benefit is allocated at the EoL.\\
Despite the AI models being trained in two different countries with differing recycling strategies, the same EoL impact method is applied to both. This is due to data limitations on electronic waste streams and the proportion of materials collected, processed and recycled. Furthermore, the datasets under consideration exclusively encompass legal processes, while the environmental ramifications of illegal e-waste disposal are not included in the analysis \cite{Ficher2025}.\\

\section*{Results}

The results are presented in two sections. First, we examine the Nvidia A100 SXM 40GB GPU as a physical product, generating intermediate results at the component and system level as illustrated in Fig. \ref{fig:GPU2FU}. Second, to assess the environmental impacts of AI-as-a-service, we scale these intermediate results according to two functional unit configurations: training BLOOM and GPT-4. Finally, we validate our results by comparing them with those of an industry report and a scientific study.

\subsection*{GPU-level Intermediate Life Cycle Impact Assessment}

\begin{figure} [H]
    \centering
    \includegraphics[width=0.75\linewidth]{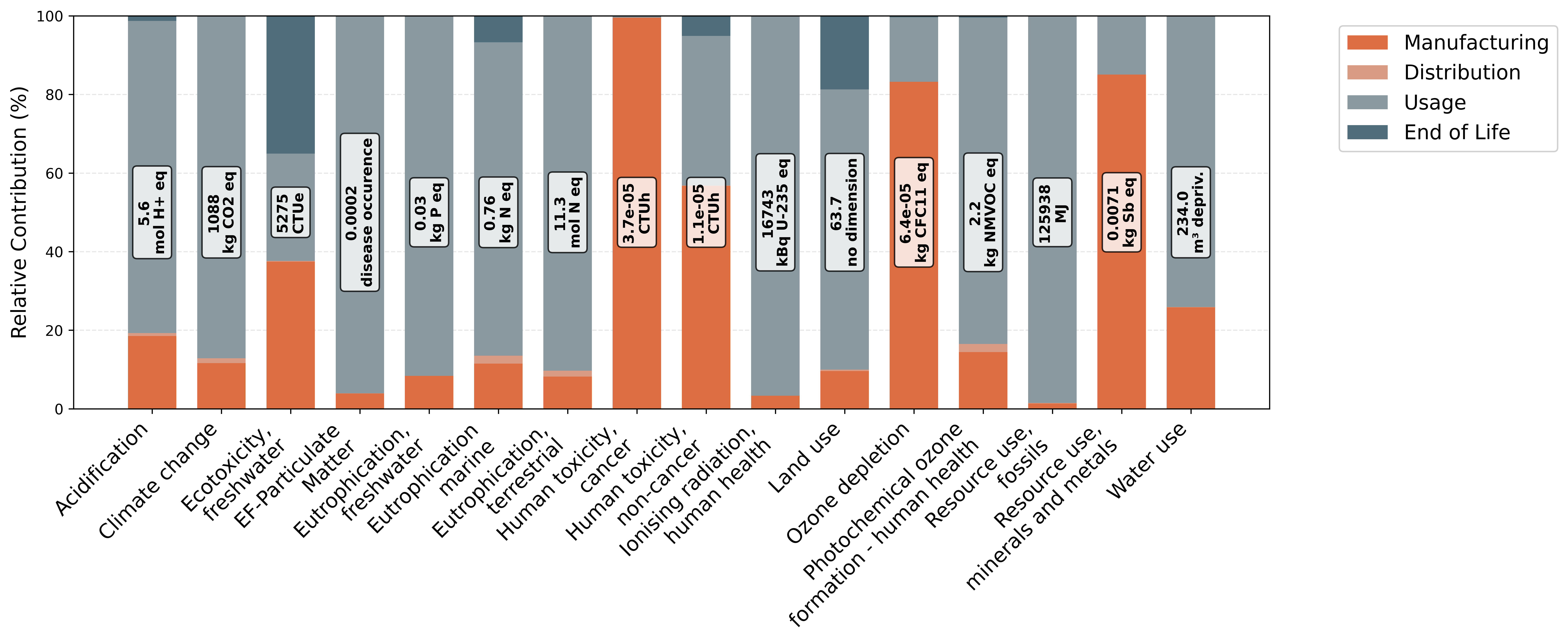}
    \caption{Product environmental impact assessment of one GPU card. Impact contribution by life cycle stage across 16 environmental impact categories. Numbers within bars indicate the total absolute environmental impact values as sum of all life cycle stages (author illustration).}
    \label{fig:product_LCA}
\end{figure}

We present intermediate impacts results at the scale of the individual Nvidia A100 SXM 40GB GPU, with detailed focus on the manufacturing stage. The assessment assumes operational parameters consistent with the BLOOM training infrastructure and context: 3-year operational lifespan, continuous operation at maximum capacity (100\% utilization) with power consumption at 400W, connected to the French electricity grid (2022 carbon intensity). 
%for one GPU card 
Environmental impacts vary at the scale of the GPU card across the four life cycle stages, i.e., manufacturing, distribution, use and end-of-life (EoL) and across 16 impact categories (see Fig. \ref{fig:product_LCA}).\\
The use phase dominates 11 out of 16 categories. Particularly for the impact category climate change, the use phase contributes to about 87\% of total impact, despite the energy consumption being based on the French electricity mix that is predominantly based on nuclear energy. The use phase similarly contributes high shares to the categories resource use (fossil fuels) ($\geq$98\%), ionizing radiation ($\geq$96\%), EF-particulate matter ($\geq$95\%), eutrophication (terrestrial) ($\geq$90\%), and acidification ($\geq$79\%) reflecting the energy-intensive nature of GPU operations. The elevated ionizing radiation impact can be attributed to the substantial contribution of nuclear energy to France's electricity grid composition.\\
\noindent
The manufacturing stage represents the second most dominant life cycle stage, with substantial contributions to 5 out of 16 impact categories. 
Specifically, human toxicity (cancer) ($\geq$99\%), human toxicity (non-cancer) ($\geq$56\%), resource use, minerals and metals ($\geq$85\%) and ecotoxicity (freshwater) ($\geq$37\%). This reflects the material extraction and processing activities associated with product fabrication, particularly the use of metals and chemical substances in manufacturing processes. 
The substantial contribution to ozone depletion ($\geq$83\%) partially reflect the release of HFC emissions and other ozone-depleting substances during semiconductor manufacturing.
Despite modeling the component transportation via air transport, the distribution contributes modestly across most categories, with its highest impact in marine eutrophication ($\geq$2\%) and terrestrial eutrophication ($\geq$1\%).
The EoL stage contributes most substantially to freshwater ecotoxicity  ($\geq$35\%) and land use ($\geq$18\%)\footnote{We exclude the analysis of water use results for the EoL stage, as the background database is subject to error in this particular criterion.}

\begin{figure} [H]
    \centering
    \includegraphics[width=0.75\linewidth]{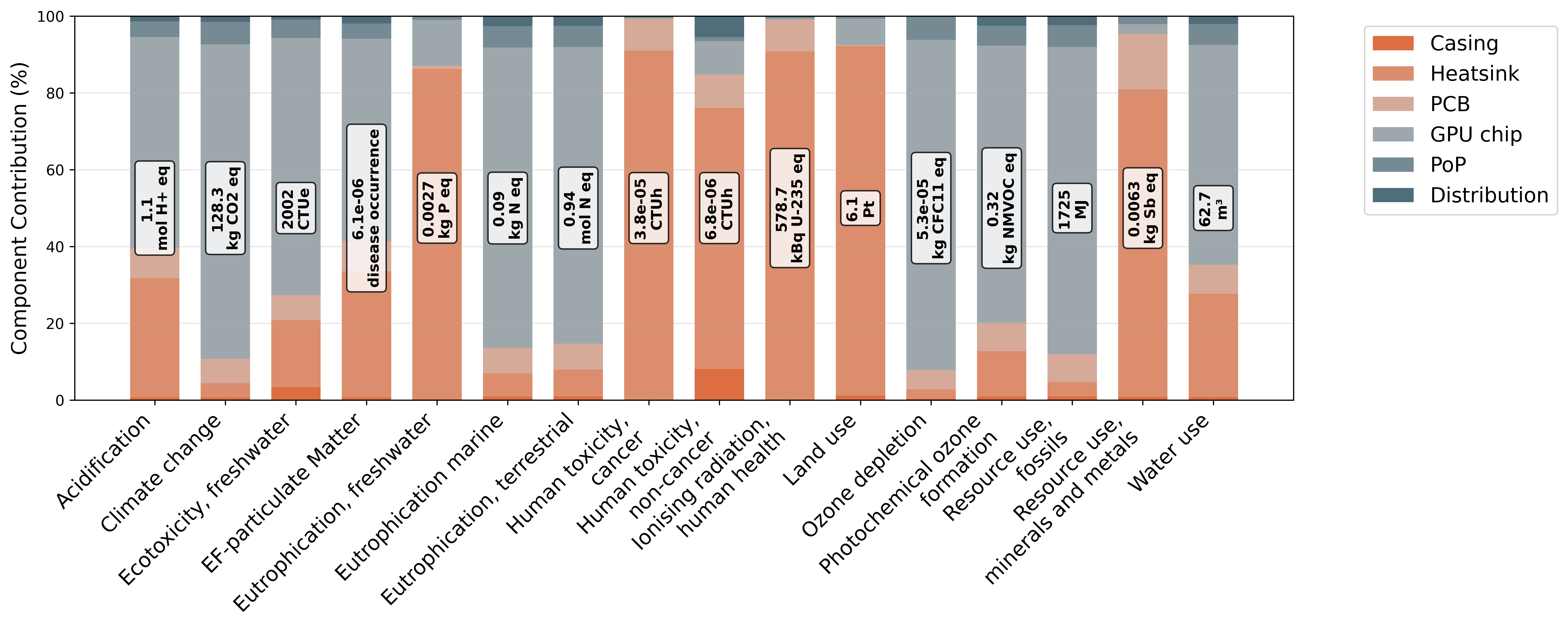}
    \caption{Environmental impact distribution across GPU component categories (casing, heatsink, PCB, GPU chip (main die + VRAM), PoP, and upstream distribution) for the cradle-to-gate stages of the Nvidia A100 GPU. Numbers within bars indicate the total absolute environmental impact values as sum of all component categories (author illustration). }
    \label{fig:by component}
\end{figure}

However, these contributions should be interpreted with caution due to significant data limitations in modeling electronic waste streams. The environmental impacts are likely underestimated, as current LCA databases do not account for illegal and informal disposal practices that are prevalent in global e-waste management \cite{Ficher2025}. Additionally, the relatively modest contribution of the EoL stage compared to manufacturing and use phase reflects the relative dominance of the other life cycle stages rather than indicating low absolute environmental impacts from disposal.\\
The results demonstrate hotspots in the product's life cycle, with the use phase representing the primary environmental burden for most impact categories, followed by manufacturing for material-intensive and toxicity impact categories and EoL for specific disposal-related categories.
%cradle-to-gate by component:
Given manufacturing's substantial contributes to the overall GPU environmental impacts, we further examine the manufacturing impacts per component across all impact category (see Fig. \ref{fig:by component}). The manufacturing stage encompasses the emissions associated with resource extractions, mineral processing, distribution and component manufacturing, collectively representing the cradle-to-gate stage. The component-level cradle-to-gate impacts assessment reveals heterogeneous environmental burden distribution across five primary elements: GPU chip, printed circuit board (PCB), heatsink, power-on-package (PoP), casing, and upstream distribution (see Fig.\ref{fig:by component}).\\
\noindent
%
%main die + VRAM
The GPU chip emerges as the dominant contributor across 10 out of 16 of impact categories and shows particularly pronounced contributions to climate change (81.8\%), and resource use, fossils (80\%), reflecting the energy intensive semiconductor fabrication processes required for GPU chip manufacturing compared to simpler ICs in the PCB. These high manufacturing impacts are further intensified by the A100's advanced 7nm process technology, where smaller node size require increasingly energy and resource-intensive manufacturing processes per unit area. The transition to finer lithography processes necessitates more complex fabrication steps, higher precision manufacturing equipment, and greater energy consumption per cm$^2$ of silicon area \cite{Pirson2023_ICproduction, Boakes2023-jm}, contributing to the relatively high environmental impact of the component. Regarding abiotic resource depletion of elements (ADPe)\footnote{ADPe reflects the use of nonrenewable resources in relation to their scarcity.}, the main die does not show high impacts because gold is primarily associated with connections on the PCB or within specific types of IC packages (bonding) on the PCB rather than the main die itself. In contrast, the heatsink's large copper content drives the overall resource depletion results for this category (80.1\%), overshadowing the gold impact present in relatively small amounts within the PCB.
%heatsink
Furthermore, the heatsink dominates human toxicity categories, contributing 91\% to cancer-related impacts and 68.0\% to non-cancer impacts. This can be attributed to the heatsink's high copper content, comprising 98\% of the component's weight. As a heavy metal, which extraction and processing can further release toxic by-products \cite{izydorczyk2021copper}, copper drives these toxicity impacts. 
Overall, the heatsink component shows impacts primarily in material-intensive categories, which is consistent with the extensive copper extraction and processing activities required for thermal management components. This component also contributes dominantly to eutrophication, freshwater (86\%), ionising radiation, human health (90.7\%), and land use (91\%).
These impacts likely stem from copper mining operations, which typically involve large-scale excavation (land use impacts), acid mine drainage contributing to surface water contamination \cite{izydorczyk2021copper} (accounting for eutrophication), and processing of ores containing naturally occurring radioactive materials \cite{chau2008naturalradiat_copper} (ionising radiation impacts).\\
%PCB
While the PCB does not dominate any impact category, it shows its largest contribution to resource use in the minerals and metals category (14.4\%), which can be explained by the gold content of the pins and connectors on the board, as well as the presence of other precious metals such as platinum, palladium, and silver. The PCB also contains toxic metals and minerals, such as arsenic, cadmium, lead and chromium.
% PoP and upstream transport 
The PoP and upstream distribution show relatively modest but consistent contributions across most impact categories, ranging from 0 to 6\%. 
These low shares for the PoP should be interpreted with caution as there are limitations in LCA databases that do not provide data representative of  the manufacturing process of such specialized components. The high integration density and complexity of PoPs, as well as the substantial electrical currents they must handle, suggest that manufacturing processes are likely underestimated (particularly for packaging and integration for processes). While the impacts of their raw material acquisition are considered to be of good quality due to the dedicated elemental analysis of the PoPs, the assessment of the manufacturing processes relies on a simplified proxy that could be underestimated. However, improving this modeling is currently very difficult due to the absence of dedicated LCA data for these components. The contribution of PoPs to manufacturing impacts should therefore be understood as a low-quality lower bound proxy with inherent uncertainties.\\
%casing
The casing component exhibits the lowest overall environmental contribution, contributing less than 1\% to most categories, indicating the relatively simple manufacturing processes required for GPU housing components.
%%%%%
The manufacturing results for one A100 reveal clear environmental hotspots at the component level. Fabrication of the main die in semiconductor facilities represent the primary environmental burden for energy and climate-related impacts, while material processing activities for the heatsink and PCB dominate toxicity and resource depletion categories. This is consistent with other studies showing that ICs often represent a large share of manufacturing impacts \cite{Pirson2023_ICproduction}.\\
To translate these intermediate product-level results into the impacts of one active GPU hour (our reference flow), we apply equation \ref{eqn:elec_conso_gpu} enabling standardized comparison across different AI training scenarios and facilitating the subsequent scaling to our FU configurations to represent the environmental impacts of complete model training processes.
The relative distribution of impact contributions per one GPU hour remains unchanged as shown in Figure \ref{fig:product_LCA}; only the absolute environmental impacts per metric will be scaled down to reflect the impacts of one active GPU hour. This hourly metric then serves as the basis for estimating the final life cycle impact assessment (LCIA) results according to equation \ref{eqn:impact_FU}.

\subsection*{AI Model Training Life Cycle Impact Assessment}

We present the LCIA results for two FU configurations, which characterize the training of GPT-4 and BLOOM. This approach captures the environmental implications of AI-as-a-service considered at the GPU cluster scale, where computational resources are deployed temporarily for specific training objectives before being reallocated or decommissioned. Subsequently, the results are normalized following the planetary boundary (PB) framework to contextualize impact magnitudes relative to Earth system thresholds. Given substantial variability in environmental outcomes across operational parameters, we conduct sensitivity analyses spanning realistic ranges of hardware utilization, lifespan, and power consumption to capture context-dependent uncertainty in impact estimates.
\begin{figure} [H]
    \centering
    \includegraphics[width=0.75\linewidth]{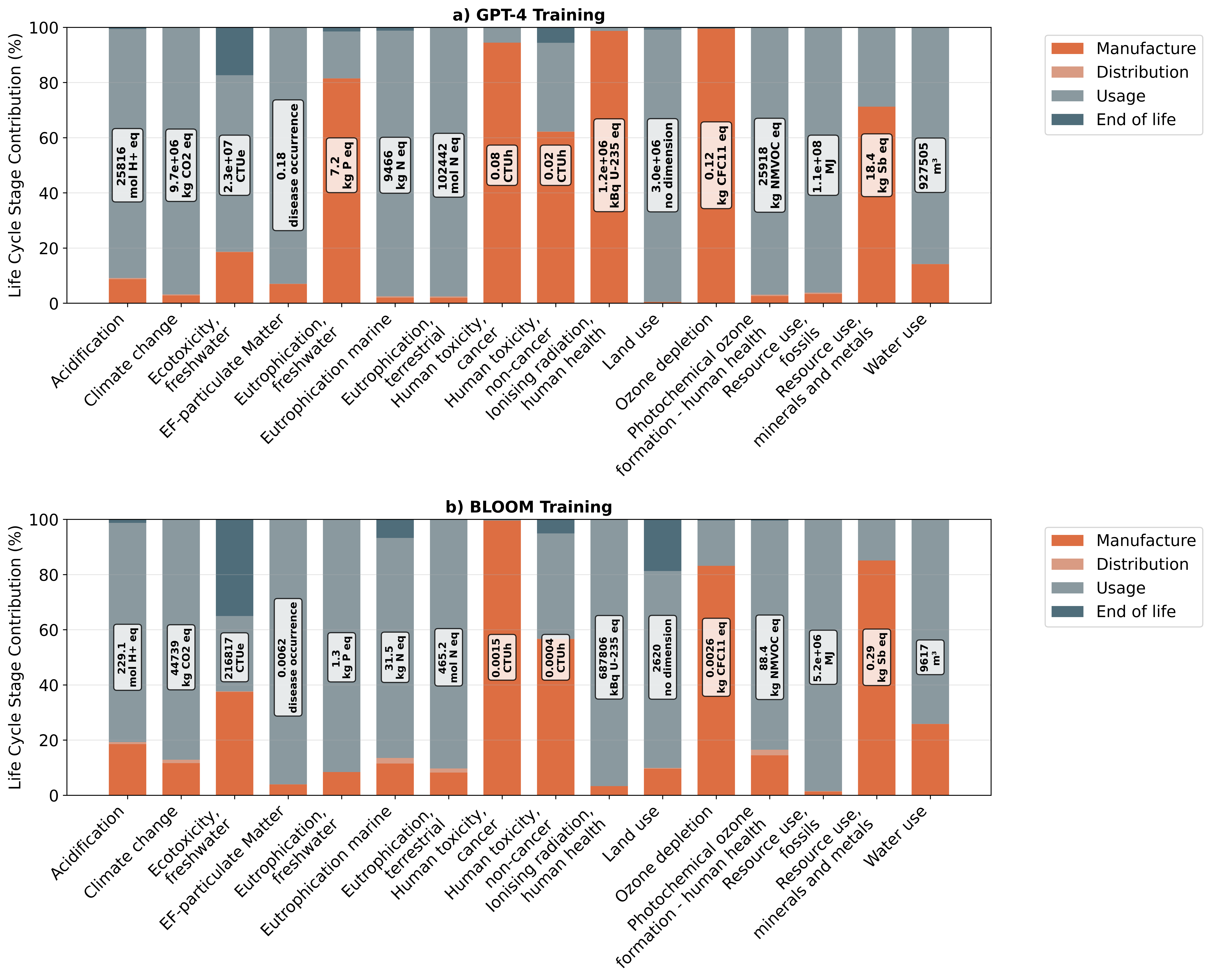}
    \caption{Impact contribution by life cycle stage for 16 environmental impact categories by AI model training BLOOM and GPT-4. Numbers within bars indicate the total absolute environmental impact values as sum of all life cycle stages (author illustration).}
    \label{fig:service_LCA}
\end{figure}

\subsubsection*{Multi-criteria Environmental Impact Assessment of AI Model Training}

For GPT-4's training process, the use phase dominates the majority of environmental impacts across 10 out of 16 impact categories. More specifically, the use phase contributes 96.8\% to the impact category climate change, 96.2\% to resource use, fossils and about 97\% to both eutrophication, terrestrial and marine. The strong impacts one these categories reflect the enormous computational requirements and energy consumption associated with training LLMs. The use phase is additionally the main contributor to acidification (90.3\%) and water use (85.8\%). The manufacturing stage represents substantial contributions to 4 of the 16 impact categories. More precisely, human toxicity, cancer (94.5\%), eutrophication, freshwater (81.4\%), and resource use, minerals and metals (71.2\%) (see Figure \ref{fig:service_LCA}). 
The distribution and EoL stages contribute relatively small amounts across all impact categories, typically representing less than 2\% each. This low contribution reflects two key factors: first, both stage's impacts are amortized over the millions of GPU hours during the operational phase; second, our model doesn't include illegal e-waste flows at the EoL, which could significantly increase EoL impacts if properly quantified.\\
The use phase dominates 11 out of 16 impact categories for training BLOOM, including climate change (86.9\%), resource use, fossils (98.5\%), and eutrophication, terrestrial (90.1\%).  The manufacturing stage accounts for substantial contributions across 5 of the 16 impact categories, more precisely resource use, minerals and metals (85.1\%), and ozone depletion (83.2\%). Further, the manufacturing stage contributes most to the toxicity related impact categories, human toxicity, cancer (99.6\%), human toxicity, non-caner (56.6\%), and freshwater ecotoxicity (37.5\%). The exact amounts can be found in the SM.

\subsubsection*{Planetary Boundaries Normalization}

Normalizing the LCIA results of training BLOOM and GPT-4 to the planetary boundaries (PBs) per capita translates the complex LCA results into an intuitive framework that communicates environmental significance in the equivalence of the global ecological limits at the scale of one human inhabitant of the earth. This makes the results more accessible to policymakers and stakeholders outside the LCA community.
To contextualize the LCA results within this framework, impacts are normalized to the PB-aware reference situation as provided by \cite{sala2018_PB_LC_development} for the year 2023 (world population of 8.01 billion \cite{WorldBank2023Pop}) corresponding to the year in which GPT-4 was trained\footnote{Although BLOOM was trained in 2022, adjusting the normalization to the 2022 world population would alter the results by less than 1\%. Consequently, the same normalization procedure is employed for both models.}.\\
\begin{figure} [H]
    \centering
    \includegraphics[width=0.75\linewidth]{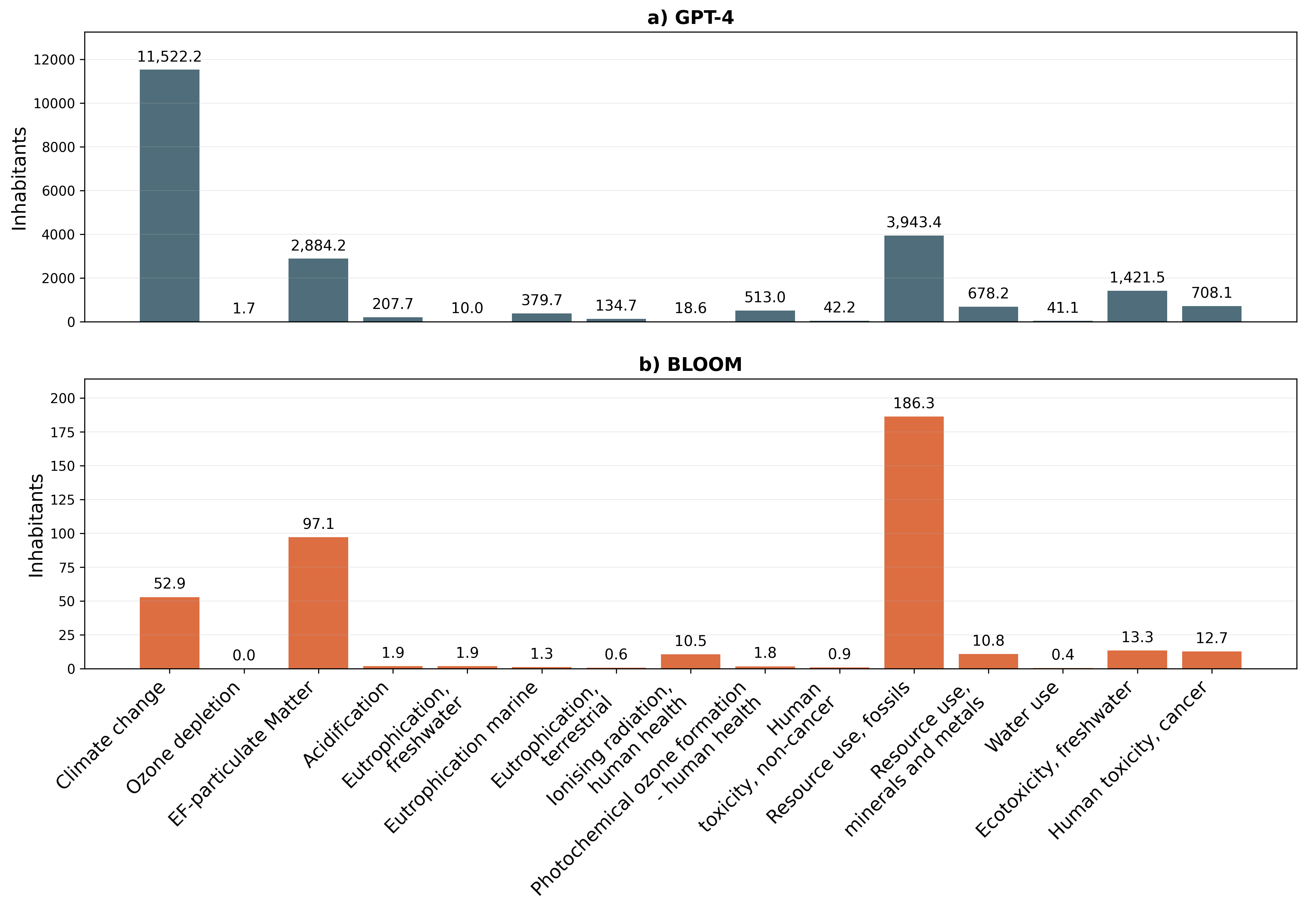}
    \caption{Life Cycle Impact Assessment relative to Planetary Boundaries for the two FU configurations: a) GPT-4 trained in the US, and separately for b) BLOOM trained in France (author illustration).}
    \label{fig:PB_LCA}
\end{figure}
\noindent
The resulting normalized metric expresses the environmental impacts of training an AI model in the equivalence of the annual `environmental budget' per capita across 15 environmental impact categories\footnote{The land use impacts are not presented, as no PB normalization factor has been developed for this LCA impact category.}.\\
%GPT-4
Training GPT-4 in Iowa is equivalent to the climate change PB budget of 11,522 inhabitants on the planet and 3,943 for the fossil resource use (see Figure \ref{fig:PB_LCA}). Those environmental impacts are primarily attributed to the carbon-intensive Iowa (US) electricity grid, which relies predominantly on fossil fuels for electricity generation. 
Additionally, the particulate matter impact (2,884), is also primarily attributed to the Iowa electricity grid, while ecotoxicity, freshwater (1,421) stems from both the electricity grid and manufacturing processes. The toxicity impacts, including human toxicity, cancer (708) and ADPe (678), reflect the complex chemical processes involved in advanced semiconductor chip manufacturing. These non-climate impacts stem primarily from the manufacturing phase of the computational hardware required for training, highlighting how the material intensity of AI hardware contributes significantly to environmental burdens beyond carbon emissions.
In total, training GPT-4 is higher than the PB-aware budget of one inhabitant of the world in all impact categories.\\
% BLOOM
Training BLOOM in France is comparable to the fossil resource use budget of nearly 200 inhabitants on earth. The fossil resource impact, despite France's low-carbon nuclear electricity grid, can be explained by two factors: the classification of uranium as a fossil fuel resource within this impact category, and the lower thermal efficiency of nuclear power plants compared to modern fossil fuel plants, which requires greater fuel input (uranium) per unit of electricity generated. Additionally, training BLOOM is nearly equivalent to the annual particulate matter budget of 100 inhabitants and is comparable to the annual climate change budget of 53 inhabitants (see Figure \ref{fig:PB_LCA}). 
11 of 15 impact categories are above the annual PB-aware budget of 100 people.\\

\subsubsection*{Sensitivity Analysis}

The GPU \textit{average power}, \textit{lifespan} and \textit{utilization ratio} are subject to significant variability depending on the deployment context, directly influencing environmental impact assessments. To capture this variability, we conducted sensitivity analyses (SA) on these three parameters using a min-max approach spanning realistic operational ranges.
The \textit{average power }consumption exerts a direct linear influence on electricity consumption (Equation \ref{eqn:elec_conso_gpu}) and depends among others on utilization intensity and cooling system design. Our reference scenario assumes an average power equivalent to the GPU's TDP of 400W. The PCIe version of the NVIDIA A100 operates at 300 W TDP, while specialized cooling configurations can sustain power draw up to 500 \cite{NvidiaA100Product}. We therefore varied this parameter between 300 W and 500 W to encompass standard and high-performance operational conditions.\\
GPU \textit{lifespan} affects the amortization of upstream impacts (raw material acquisition and manufacturing) and downstream end-of-life burdens across total GPU hours (Equation \ref{eqn:gpu_per_hour}). Hardware durability depends on several factors, including utilization intensity and thermal management effectiveness. Our reference scenario assumes a 3-year lifespan, with sensitivity analysis spanning one to four years to reflect variability in replacement cycles across different operational contexts.\\
The \textit{utilization ratio} influences both \textit{electricity consumption} (Equation \ref{eqn:elec_conso_gpu}) and impact amortization over active \textit{GPU hours} (equation \ref{eqn:gpu_per_hour}). Our reference scenario assumes 100\% utilization, representing a maximalist approach, with sensitivity analysis ranging from 50\% to 85\% to reflect realistic operational variability.
An increase, or a decrease, in power consumption is accompanied by a corresponding increase, or decrease, in the magnitude of impacts. This is particularly noteworthy for criteria that are strongly influenced by the use phase, such as climate change, particulate matter and the depletion of fossil resources, which vary between 75\% and 125\%. These findings underscore the importance of energy efficiency interventions for mitigating grid-related environmental impacts, with effectiveness contingent on data center location and electricity supply characteristics.\\
For manufacturing phase dominated impact categories, both \textit{lifespan} and \textit{utilization ratio} exert substantial influence on results.
Human cancer toxicity exhibits the greatest variability, reaching approximately 300\% for both GPT-4 and BLOOM, while ionizing radiation for GPT-4 varies from nearly 300\% at one-year lifespan to 75\% at four years.
However, these variations reflect cluster-level impacts rather than individual card-level dynamics.
Extended \textit{lifespan} reduces replacement frequency, thereby decreasing the number of newly manufactured cards and their associated environmental burdens. Similarly increased \textit{utilization ratio}, reduces at the cluster level the total number of cards required to deliver equivalent computational capacity through more efficient resource sharing, with clusters providing more GPU hours from the same hardware inventory by minimizing idle time. 
Both mechanisms ultimately reduce aggregate hardware production and its embodied impacts. These findings demonstrate that maximizing utilization through workload optimization and hardware sharing, coupled with extended replacement cycles, represents a crucial strategy for mitigating human, non-human and ecological toxicity impacts alongside mineral and metal resource depletion associated with AI infrastructure.
\begin{figure}
    \centering
    \includegraphics[width=1\linewidth]{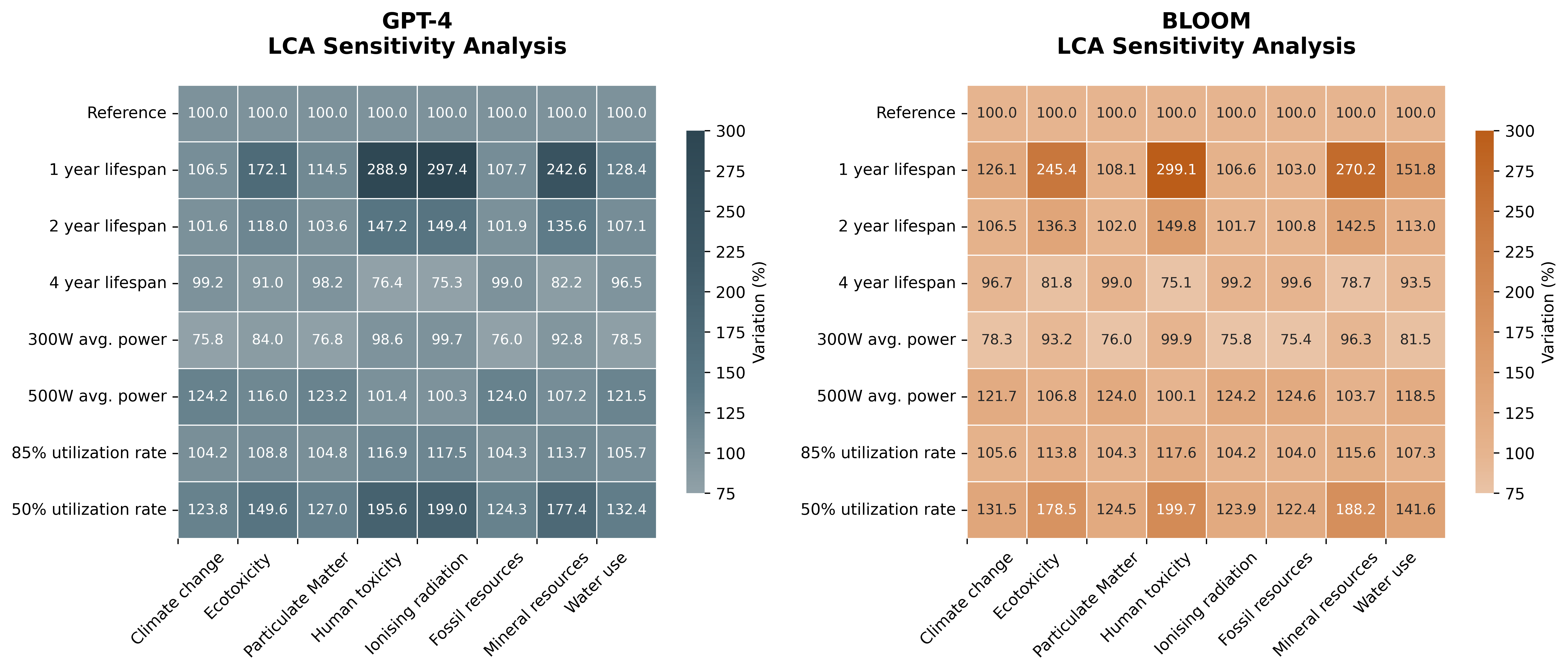}
    \caption{Sensitivity analysis for GPT-4 and BLOOM. The reference scenario is defined with a 3-year lifespan, 400 W average power consumption and 100\% utilization rate (author illustration).}
    \label{fig:SA-heatmap}
\end{figure}
\noindent

\subsubsection*{Trade-off between lifespan and utilization ratio}
The three variables considered in the SA are not independent. One critical trade-off influencing the environmental impacts of AI training is between maximizing GPU utilization and preserving hardware longevity. Professional data center environments (including training cluster divisions) can optimize GPU operation to maximize computational efficiency and return on investments. A higher GPU utilization rate generally leads to lower per-task environmental impact, as the embodied emissions are spread over more useful work. In contrast, lower utilization amplifies the significance of manufacturing impacts in the LCA results. \\
However, higher utilization rates create sustained thermal stress that accelerate silicon degradation. This can manifest as decreased performance, increased latency, and a higher susceptibility to hardware errors over time \cite{karydopoulos2025heatimpact}. In addition, high utilization - continuous near-maximum power draw - generates substantial heat loads that contribute to various failure mechanisms. These effects compound over time \cite{Titan2020}, potentially reducing the effective operational lifespan of the hardware and thereby increasing the amortized environmental impacts per unit of computational work performed.
Even systems with modest aggregated utilization rates may experience prolonged periods of maximum power draw during training phases, subjecting the hardware to the same thermal stress as continuously high-utilization rates. This suggests that simply reducing allocation rates may not necessarily extend hardware lifespan or reduce life cycle environmental impacts. The reason is that AI training workloads are characterized by intensive computational bursts that push GPUs to their thermal limits regardless of overall utilization patters. During these training phases, GPUs operate at peak power consumption for extended periods, generating the same thermal stress and degradation mechanisms as would occur under continuous high utilization.\\
The interdependence between power consumption, thermal stress, and hardware lifespan suggests that optimization strategies should consider the full system dynamics rather than isolated metrics. However, because GPUs represent a significant capital investment, maximizing their allocation utilization is often a top priority for operators seeking efficient return on hardware expenditures. Yet, the economic optimization models used in practice do not account for the environmental costs of accelerated hardware replacement cycles. AI model developers and system architects can minimize the per-unit embodied environmental impact by ensuring maximum operational efficiency of system resources and extending the service life of AI infrastructure \cite{wu2022sustainable}.

\subsection*{Validation Against Existing Studies}

We discuss the consistency of our findings with previous estimates of GPU embodied impacts and AI training emissions through two studies: Nvidia's HGX H100 carbon footprint report \cite{NvidiaH100PCF} and the life cycle impacts of BLOOM's model training \cite{luccioni2023}. Given the absence of reproducible multi-criteria assessments of GPUs, these comparisons consider exclusively the climate change impact category.

\subsubsection*{Comparison with industry report: Nvidia's HGX H100 System Product Carbon Footprint Report}

Nvidia's 2025 cradle-to-gate product carbon footprint (PCF) for the HGX H100 system reports 1,312 kg CO$_2$eq \cite{NvidiaH100PCF}. The assessment focuses exclusively on raw material extraction, mineral processing and component manufacturing for the climate change category only, enabling comparison with our intermediate LCIA results at the component level for the same impact category. The primary contributors to the carbon emissions are the materials and components, which represent 91\% of the total footprint, while the assembly process accounts for an additional 8.6\% and transportation contributes minimally at 0.4\% to the overall footprint. For the materials and components category, the report provides a detailed breakdown by component type. High bandwidth memory emerges as the largest contributor (42\%), followed by ICs (25\%) and thermal components (18\%). Nvidia also identifies electromechanical components (4\%), common components (0.8\%), PCBs (0.7\%), mechanical components (0.6\%), and interconnects (0.4\%) as contributing to the overall PCF.\\
Direct comparison between the Nvidia study and this study presents several methodological challenges. 
The Nvidia assessment analyzes the complete HGX H100 system comprising 8 GPUs, baseboard, networking infrastructure and cooling components, whereas our study examines a single GPU with detailed component-level breakdown. Additionally, the H100 uses 5nm lithography technology while the A100 uses 7nm technology, resulting in different manufacturing processes and material requirements. 
Despite these differences, both studies encompass the cradle-to-gate stage and employ kg CO$_2$eq. as metric, making cautious analysis of analogous component categories feasible with appropriate consideration of these methodological limitations.\\
Our cradle-to-gate analysis of the Nvidia A100 GPU estimates 127.6 kg CO$_2$eq per GPU, dominated by the GPU chip and VRAM (82.3\%), followed by the PCB (6.4\%) and heatsink (3.8\%). When grouped analogously, semiconductor components (main die, VRAM, PoP) represent 88.7\% of the total, aligning closely with Nvidia's 67\% for memory and ICs\footnote{While the PoP is not technically an IC, it is largely comparable to the components listed as `active components' in the Nvidia report.}.
Our PCB analysis shows 6.4\% contribution, which is  5.7\% higher than Nvidia's assessed PCB contribution to the overall carbon footprint. The substantial difference in system perimeter - GPU PCB in this study versus the HGX baseboard PCB in the Nvidia report - makes direct comparison uncertain. The thermal management components (heatsink) in our analysis contributes 3.8\% of the carbon product footprint, approximately 15\% lower than the assessment in the Nvidia report.\\
Despite architectural and methodological differences, most notably system scope (single GPU vs. multi-GPU system), lithography (7 nm vs. 5 nm), and database choice (Negaoctet vs. ecoinvent/Sphera), the proportional distribution of emissions across component categories shows strong alignment between the two assessments and both studies identify semiconductor components (specifically integrated circuits) as the dominant carbon hotspot in GPU production. 
\begin{wrapfigure}{l}{0.5\textwidth}
    \centering
    \includegraphics[width=0.52\textwidth]{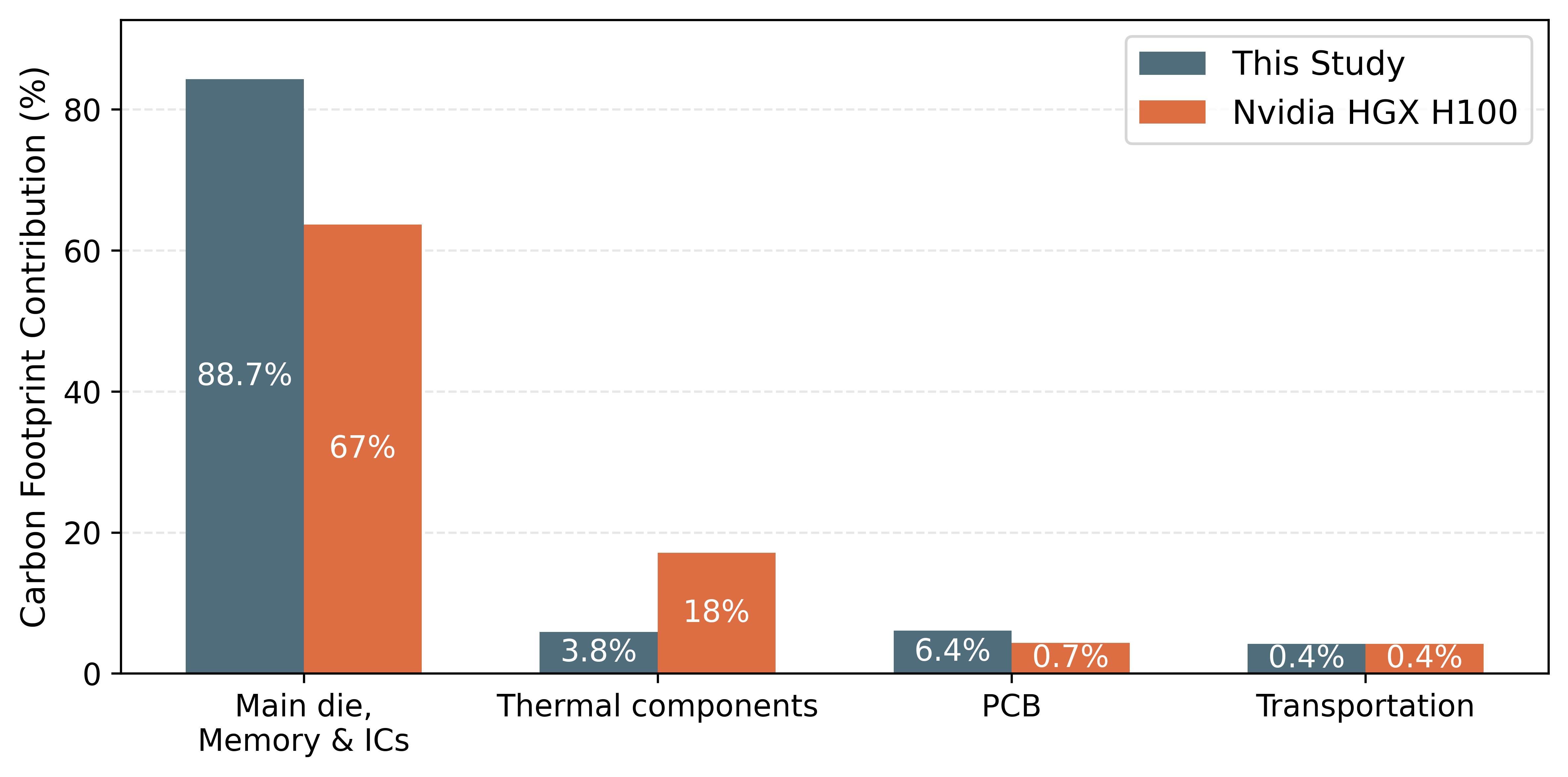}
    \caption{Comparison of contribution to total product carbon footprint (\%) by component group estimated by this study and the Nvidia PCF report for the HGX H100 (author illustration).}
    \label{fig:NvidiaHGXH100_ThisStudy}
\end{wrapfigure}
\noindent
Nevertheless, notable discrepancies emerge when accounting for memory capacity differences. The H100 incorporates 80GB of high bandwidth memory, which is double compared to our assessed GPU's specifications, yet the observed emissions differential exceeds what memory scaling alone would predict. For one, this discrepancy might stem from methodological variations, particularly regarding data sources. Nvidia utilized ecoinvent 3.10 and Sphera’s LCA databases, whereas our assessment employed the Negaoctet database. Each database incorporates different modeling assumptions for semiconductor manufacturing processes. 
For the climate change category, there may be less variation between databases compared to other environmental indicators, although differences in system models and elementary flow can still contribute to discrepancies. Despite these methodological differences, both studies converge on identifying semiconductor components (specifically integrated circuits) as the primary carbon hotspot in GPU manufacturing, whether examined at the individual GPU scale or the server system level.

\subsubsection*{Comparison with scientific literature: LCA Results of Training BLOOM}

Luccioni et al. (2023) estimated the climate change impacts of training BLOOM encompassing the impact of the GPU cards during both the training and inference phases, assuming an embodied footprint of 150 kg CO$_2$eq per A100 GPU based on secondary data \cite{luccioni2023}. 
\begin{wrapfigure}{r}{0.5\textwidth}
    \centering
    \includegraphics[width=0.40\textwidth]{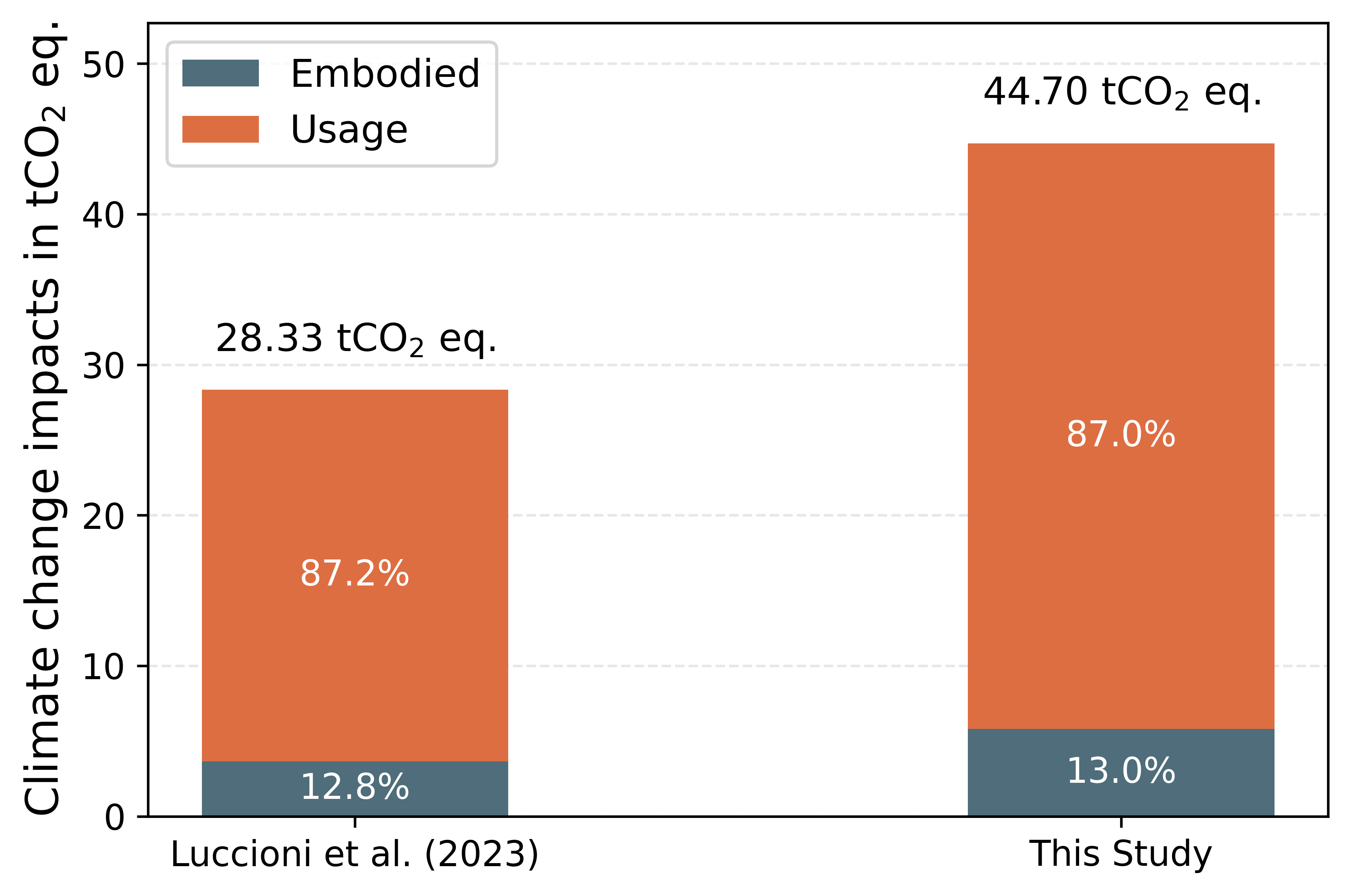}
    \caption{Comparison of total emission results in tCO$_2$ eq. estimated by Luccioni et al (2023) \cite{luccioni2023} and by this study (author illustration).}
    \label{fig:bloom_vs_this_study}
\end{wrapfigure}
\noindent
To ensure equivalent system boundaries, the following comparison focuses solely on the embodied and usage impacts of the GPU cards themselves in the context of AI training. Our cradle-to-grave analysis yields a comparable value of 141 kg CO$_2$eq per GPU. It should be noted that our estimate includes the impacts of distribution (12\%) and disposal (1,5\%), which are not included by Luccioni et al. (2023), suggesting the previous estimate is conservative. With a lifespan of six years and 85\% utilization ratio compared to our three-year lifespan and 100\% utilization, the embodied impacts per GPU hour diverge: from 0.003 kg CO$_2$eq in the original study to 0.0054 kg CO$_2$eq in this study. However, under consistent lifetime and utilization assumptions, both studies converge on an embodied impact of approximately 0.003 kg CO$_2$eq per GPU hour. Their global estimate for the embodied impacts of training BLOOM is 3.64 t CO$_2$eq, compared to our estimate of 5.8 t CO$_2$eq.\\
For the use phase, Luccioni et al. report 24.7 t CO$_2$eq compared to 38.9 t CO$_2$eq in this study, primarily due to differences in the electricity carbon intensity applied (57 g CO$_2$/kWh vs. 90 g CO$_2$/kWh) (see Fig. \ref{fig:bloom_vs_this_study}). The difference likely stem from Luccioni et al. employing a production mix that does not account for transmission and distribution losses in the network as well as the import and export of the electricity \cite{FRcarbonIntensity_source} while the CODDE database employed in this study accounts for these factors.\\
Despite methodological variations, results align closely in both absolute value and distribution across life cycle stages, validating the robustness of prior carbon-based assessments. Hence, our primary data collection approach validates that carbon-related estimates in previous studies are reasonably accurate. This is consistent with the fact that carbon accounting results exhibit less variation between studies due to standardized methodologies and relatively consistent characterization of elementary flows related to climate change across LCA databases.
However, integrating primary teardown and elemental data reveals that non-carbon impact categories differ more substantially. Adjusting the life cycle inventory based on our compositional analysis increased total impacts by an average of 3.9\%, with up to 32.8\% for abiotic resource depletion of minerals and metals.\\
This underscores that while carbon estimates remain relatively stable across studies (using secondary versus primary data), other environmental impact categories are more sensitive to primary material data, emphasizing the need for primary data collection and comprehensive, multi-criteria assessments of AI hardware.
\subsection*{Discussion: Beyond Carbon Tunnel Vision - Multi-Criteria Environmental Assessment of AI}

This study addresses a critical gap in current AI sustainability discourse by examining environmental impacts beyond carbon emissions that currently dominate both academic discourse and public debate.
Our cradle-to-grave LCA of AI model training examines 16 environmental impact categories, representing the first comprehensive multi-criteria environmental evaluation of large-scale AI systems.
By systematically quantifying impacts ranging from resource depletion and ecotoxicity to human health effects, we demonstrate that `carbon tunnel vision' fails to capture the full spectrum of environmental consequences associated with AI development. However, not all 16 impact categories can be interpreted equally. Most impact categories in LCA treat environmental burdens as spatially uniform or at a very low level of spatial precision, with the notable exception of water use, which includes location-specific characterization factors during the impact assessment phase. This limited spatial differentiations between local and global impacts complicates direct comparisons across categories and obscures targeted action to address environmental concerns.
More fundamentally, the impact category climate change is a global concern with well-understood mechanisms, while others (like human toxicity or ecotoxicity) manifest as highly localized impacts and are less intuitive for most stakeholders to interpret, and require more environmental research to properly understand the mechanisms behind the environmental impacts pathways.\\ 
The case of GPT-4 training exemplifies these interpretative challenges through striking disparities across life cycle stages and impact categories. GPU hardware manufacturing accounts for 94.5\% of human toxicity (cancer) impacts yet contributes less than 3\% to total climate change impacts. In contrast, the operational phase contributes almost 97\% to the climate change category but only 5.5\% to human toxicity (cancer) (see Fig. \ref{fig:service_LCA}). 
The abstract unit `Comparative Toxic Units for humans' (CTUh) used to measure the impact category human toxicity (cancer) does not convey the scale or distribution of health outcomes, which complicates interpretation efforts.
For all non-LCA practitioners; the indicator CTUh expresses the potential carcinogenic risk increase in humans from exposure to carcinogenic substances, representing the estimated increase in morbidity cases in the total human population per unit mass of a substance emitted. However, this cryptic unit makes it difficult to quantify the scale of the underlying environmental impacts, limiting stakeholder comprehension and policy relevance.
More critically, since LCA results are not site-specific, it remains nearly impossible to interpret how these cancer impacts actually affect specific populations or determine which exact populations bear these health risks. While manufacturing-related carcinogens likely affect nearby communities, and/or communities near extraction sites, the CTUh units does not indicate whether this affects one person severely or one million people minimally.\\
Recent research has begun to address this spatial blindness by identifying where life cycle stages take place, also called `regionalized LCA'.
For example, component manufacturing-related health impacts are geographically concentrated near semiconductor facilities, which are primarily located in East and Southeast Asia \cite{roussilhe2022, wang2023semicon_water}, where environmental health protections may be less stringent than in region where AI training occurs. 
Water consumption presents a particularly acute example: semiconductor manufacturing is both water-intensive and geographically concentrated in water-stressed regions experiencing recurrent drought conditions, including Taiwan and other key production hubs \cite{roussilhe2022, TaiwanDrought_Feng2023}. 
These spatial patterns reveal a fundamental asymmetry in the distribution of AI's environmental costs and benefits. Populations bearing the health and resource burdens of the hardware life cycle are often not the same as those able to access AI services \cite{falk2024attribution}, highlighting inequalities throughout the value chain that remain invisible in conventional environmental assessments.\\ 
This geographic decoupling of impacts and benefits raises critical environmental justice concerns that remain obscured in spatially aggregated assessments, underscoring the need for attribution frameworks that can trace environmental responsibility across global AI supply chains and inform equitable governance mechanisms.
An environmental justice lens highlights the linkages between environmental and social conditions and sheds light on how different levels of environmental quality and protection contribute to the health and well-being of some groups, while harming the welfare of others \cite{EJ_OECD2024}. 
 In the context of AI development, environmental justice would then encompass the equitable distribution of environmental benefits and burdens, ensuring no population bears disproportionate negative consequences from AI hardware production and use while others capture technological and economic gains. However, current research indicates substantial inequities across AI supply chains: populations in Western and technologically advanced economies, the primary AI service beneficiaries, experience predominantly diffuse global impacts, particularly climate change, along with localized effects such as particulate matter depending on electricity infrastructure. Conversely, marginalized populations in extraction, processing, manufacturing, and disposal regions receive minimal service benefits or access to AI systems while bearing both concentrated local environmental degradation (toxic emissions, resource depletion, ecosystem damage) as well as global climate impacts \cite{zhou2022carbon_Inequal., falk2024attribution}.
The resulting impact profile suggests systematic environmental injustice embedded within AI development pathways, where global supply chain configurations externalize environmental costs to communities with limited regulatory protection or political voice creating disproportionate risks for vulnerable populations \cite{EJ_OECD2024, mensah2025missing}.
The geographic disconnect between impact and benefit illustrate how LCA methodologies are limited to capture global inequalities embedded in the AI supply chain, because elementary flows are typically treated as global rather than spatially differentiated.
Additionally, the interpretation of non-climate impact categories faces challenges by mixing midpoint and endpoint assessments creating methodological inconsistencies. Placing human health impacts (aggregated endpoint categories) on the same analytical level as e.g. water consumption (midpoint) lacks strict methodological rigor. 
Nonetheless, the solution is not to neglect non-carbon impacts, but to develop tools to better account for them.
Future research could explore social LCAs that could challenge the common narrative of `AI for Good' or the use of `AI for Sustainability' \cite{falk2024challenging} by revealing how AI development creates environmental winners and losers.\\ 
Carbon-only accounting fundamentally misrepresents AI's globally distributed impact profile, creating dangerous blind spots in the sustainability discourse. 
Moving beyond carbon tunnel vision requires not only expanding the scope of environmental indicators but also developing methodologies that can meaningfully address environmental justice concerns across global AI supply chains. Comprehensive environmental assessment frameworks must incorporate multi-criteria optimization approaches that account for geographic, temporal, and impact-specific distributions of environmental burdens. 
Spatially explicit toxicity and ecosystem impact assessments should become foundational to sustainable AI development strategies, contingent on critical improvements in LCA databases and characterization methods.

\section*{Limitations and Transparency}

It should be noted that this paper focuses on the use of an A100 GPU for AI training purposes. This process is characterized by substantial energy requirements and a high utilization ratio.
In the context of inference, it is probable that the embodied impacts will take a more significant importance in the life cycle of the cards. Further, our LCA analysis was based on A100 40GB GPU specifications, though the evaluated AI models were trained on both 40GB and 80GB configurations.\\
Regarding the \textit{usage impact} the model is highly sensitive to GPU \textit{lifespan}, their \textit{power consumption} and \textit{utilization ratio}. 
However, obtaining precise values for these parameters is constrained by both data limitations and their variability across different usage contexts.
In other words, they are uncertain, making them important parameters \cite{Rosenbaum2018} to further investigate. Consequently, it is advisable to consider bounds for the lifespan, maximum power draw and utilization ratio (see SM).
In the context of empirical research conducted on an equivalent perimeter with primary usage data, it is imperative to prioritize the collection of these data to ensure optimal quality.\\
Regarding the \textit{embodied impact}, generating an inventory (LCI) partly based on primary data enables us to obtain a more accurate representation of both our FU configurations. Correcting the elementary flows (specifically, metal flows) associated with the process units in the CODDE database used for modelling provides a more accurate aggregated LCI for these flows. Thus, we are confident that we also better represent the aggregated elementary flows at the scale of our FU.\\
However, a comparison of the cradle-to-gate impact estimates obtained using another background database (imec net zero,  \cite{Boakes2023-jm}) exclusively for the GPU chip (main die and VRAM)  highlights significantly different impact results (around 50\%) for the climate change criteria. This difference can be explained by modeling assumptions and by differences in the data values provided by the two data set providers, imec net zero and Negaoctet. Therefore, the results cannot be considered as ground truth, but rather depend on the background databases used \cite{su12239948}.
\section*{Conclusion}

This study provides the most advanced multi-criteria environmental data to date on a graphic card (Nvidia SXM A100), a key component of artificial intelligence. The extensive primary data collection allows for the confirmation of previously estimated carbon footprints based on proxy and secondary data (approximately 150 kgCO2eq. per card). This enhanced the reliability of literature relating to the carbon footprint of AI. Moreover, the LCA results demonstrate the value of primary data collection through detailed teardown analysis, BOM documentation and elemental composition analysis in comparison to generic data, particularly in the context of multi-criteria analysis. The findings of our assessment indicate that, while carbon footprint estimates demonstrate stability with only a 1.77\% increase when employing primary versus secondary data, other impact categories exhibit substantial variations. Particularly, the environmental impact from resource use, minerals and metals increased by about 33\%. This finding validates existing carbon estimates while highlighting the critical importance of primary data for comprehensive environmental assessments beyond climate change.\\
Secondly, by conducting a multi-criteria cradle-to-grave analysis of BLOOM and GPT-4 training, this study extends the AI sustainability discourse demonstrating that current AI environmental assessments, dominated by carbon-centric and operational emissions, underestimate the full environmental impacts of AI systems. While the use phase was already known to be an important climate change concern other impacts relate to electricity consumption should also be considered depending on the location of the infrastructure. These include health-related impacts (e.g. particulate matter, ionising radiation) and the depletion of fossil resources. Furthermore, impacts arising from the manufacturing stage are significant, including depletion of abiotic resources and impacts on human health. Finally, even if the impacts related to the operational end-of-life of the cards remains uncertain, it should also be considered in particular for eco-toxicity.\\
By advancing LCA methodology through high-quality primary data collection and expanding the Sustainable AI discourse beyond `carbon tunnel vision', this research establishes a foundation for more comprehensive AI sustainability assessments. All primary data are made publicly available in our supplementary material to support further research and enable assessments of other AI training scenarios.\\
The transition towards sustainable AI development demands fundamental changes in how we assess, communicate and govern the environmental impacts of AI, along with improved transparency from industry and developers. Future research should extend the scope of this study by including the complete AI infrastructure (cooling, networks, datacenter building,etc.) and the complete AI development life cycle (R\&D, pre-training, inference, fine-tuning, and more). These efforts should be directed towards more effectively capturing the implications of environmental justice. For instance, the development of site-specific impact assessment methodologies is essential for meaningfully addressing the geographical disconnect between AI benefits and environmental burdens.
\section*{Supplementary Material}

The Supplementary Materials (SM) contain comprehensive documentation of our primary data collection methodology and results. This includes the detailed bill of materials (BOM) inventory for all GPU components, step-by-step teardown analysis of the GPU chip and PoPs, and complete results from the multi-element composition analysis conducted on each component group. 
The elemental analysis data provides precise quantification of material composition across the main die and VRAM, Power-on-Package, printed circuit board, and heatsink, enabling full reproducibility of our LCA calculations.
To ensure complete transparency and enable future research applications, all primary datasets, analytical procedures, and component specifications will be made openly available.\\
You can find the SM here: \href{https://github.com/sophia-falk/more-than-carbon}{github.com/sophia-falk/more-than-carbon}.
\section*{Acknowledgments \& Funding}

The authors would like to thank ADEME for their material and financial support as well as the cloud providers who supplied broken GPUs for this study.
We also extend thanks to Julien Comel, Head of Research at Terra Nova Développement, who conducted the elemental composition analysis.\\
Funded by the TRA Sustainable Futures (University of Bonn) as part of the Excellence Strategy of the federal and state governments.\\
The work of Thibault Pirson in this study was supported by the SOIL project funded by the Chips Joint Undertaking under the grant 101139785.\\
Funding for this research was provided by the Alexander von Humboldt Foundation in the framework of the Alexander von Humboldt Professorship for Artificial Intelligence and endowed by the Federal Ministry of Research to Prof. Dr. Aimee van Wynsberghe.\\

\pagebreak
\printbibliography

\end{document}